\documentclass[%
 reprint,
amsmath,
amssymb,
aps,
floatfix,
superscriptaddress,
]{revtex4-2}

\usepackage{graphicx}
\usepackage{dcolumn}
\usepackage{bm}
\usepackage{nccmath}
\usepackage{amsmath}
\usepackage{amsfonts}
\usepackage{bm}
\usepackage{empheq}
\usepackage{dcolumn}
\usepackage{mathtools}
\usepackage{tensor}
\usepackage{bm}
\usepackage{subfigure}
\usepackage{comment}
\usepackage{orcidlink}
\usepackage{braket}
\begin{document}

\preprint{APS/123-QED}

\title{Response of a classical mesoscopic oscillator to a two-level quantum system}%



\author{Felipe Sobrero \orcidlink{0009-0009-6497-6612}}
\affiliation{%
 Centro Brasileiro de Pesquisas Físicas (CBPF), Rua Dr.~Xavier Sigaud 150, CEP 22290-180, Rio de Janeiro, RJ, Brazil \\
}

\author{Luca Abrahão \orcidlink{0009-0005-6266-5192}}%
\affiliation{%
Pontif\'{\i}cia Universidade Cat\'olica do Rio de Janeiro, 22451-900 Rio de Janeiro, RJ, Brazil\\
}%

\author{Thiago Guerreiro\orcidlink{0000-0001-5055-8481}}%
\email{thguerreiro@gmail.com}
\affiliation{%
Pontif\'{\i}cia Universidade Cat\'olica do Rio de Janeiro, 22451-900 Rio de Janeiro, RJ, Brazil\\
}%

\author{Pedro V. Paraguassú \orcidlink{0000-0003-2334-5688}}%
\email{venturaskate@gmail.com}
\affiliation{%
Pontif\'{\i}cia Universidade Cat\'olica do Rio de Janeiro, 22451-900 Rio de Janeiro, RJ, Brazil\\
}%

\begin{abstract}
We investigate the dynamics of a classical mechanical oscillator coupled to the simplest quantum system, a single qubit. Using the Feynman-Vernon influence functional formalism, we show that the qubit's influence manifests as both deterministic and stochastic forces on the oscillator. These forces are highly dependent on the qubit's initial quantum state, imprinting unique measurable signatures onto the oscillator's response.
The present results provide a direct pathway to quantum state reconstruction through classical noise spectroscopy. By employing the Fisher Information Matrix, we quantify the efficacy of estimating the initial qubit state from the continuous classical record, revealing a fundamental temporal asymmetry between population and phase estimation. This framework has potential applications to mesoscopic optomechanical experiments, quantum metrology, and tabletop tests of the quantum nature of gravity.
\end{abstract}


\maketitle

\section{Introduction}

Mesoscopic mechanical oscillators, such as trapped nanoparticles \cite{millen2020optomechanics, gonzalez2021levitodynamics, dania2025high, magrini2021real} and nanomechanical resonators ~\cite{aspelmeyer2014cavity, chan2011laser, cryer2025enhanced, bild2023schrodinger},
present ideal platforms to explore fundamental physics \cite{ulbricht2021testing, moore2021searching, carney2021mechanical, carney2023searches}, the quantum-classical boundary \cite{zurek2003decoherence, schlosshauer2004decoherence, anglin1997deconstructing} and the interface between quantum theory and gravity \cite{chiao2006interface, chiao2004conceptual, bose2017spin, marletto2017gravitationally, carney2019tabletop, bose2025massive, feldman2025trapping, spin_collab}. While directly preparing and observing fully quantum mesoscopic mechanical oscillators is an outstanding experimental challenge \cite{aspelmeyer2022zeh}, interesting phenomena can already be accessed and investigated by interfacing quantum and classical systems, which might provide an intermediate, easier experimental step. For instance, small-scale mesoscopic systems display interesting stochastic thermodynamical effects ~\cite{peliti2021stochastic, ciliberto2017experiments, seifert2012stochastic, strasberg2022quantum, bettmann2025quantum, jarzynskireview2011}, while the interaction of a classical oscillator to an unobserved quantum system may exhibit fundamentally new, quantum-induced statistical effects which provide a glimpse of quantum behavior \cite{paraguassu2024quantum, paraguassu2025apparent}.  Moreover, models mixing quantum-classical degrees of freedom have recently gained prominence as candidates of post-quantum theories of gravity \cite{oppenheim2023postquantum,layton2024-1, layton2024-2}. Hybrid quantum-classical systems thus provide an interesting frontier, both theoretically and experimentally. 


The hybrid quantum-classical interface can be explored in a number of distinct systems~\cite{treutlein2014hybrid}, from superconducting circuits coupled to mechanical elements~\cite{armour2002entanglement, oconnell2010quantum}, to trapped ions and levitated particles in optical and electrostatic traps~\cite{bykov2024nanoparticle, delic2020cooling, piotrowski2023simultaneous}. In this work, we focus on levitated mechanical oscillators as platforms for studying the interplay between quantum and classical systems, and investigate the interaction between a qubit and a mesoscopic mechanical oscillator. To do so, we consider the Jaynes-Cummings (JC) interaction \cite{jaynes2005comparison}, of fundamental importance to cavity quantum electrodynamics \cite{haroche2013controlling, haroche2020cavity}, quantum measurement and information  \cite{brune1992manipulation, knight2024enduring} and the quantum-to-classical transition \cite{brune1996observing}. 

{In order to eliminate the qubit degrees of freedom, we proceed neither via operator formalism nor master equation approaches (such as the Nakajima-Zwanzig projection method \cite{breuer2002theory}). Instead, we employ the Feynman-Vernon influence functional method \cite{feynman1963theory}, a powerful framework for open quantum systems \cite{hu1992quantum,calzetta1988nonequilibrium}. 
Although less frequently applied in this context, the Feynman-Vernon formalism not only provides a natural framework for perturbation theory but also circumvents some of the challenges that appear in other approaches. While standard approaches based on the Heisenberg picture primarily yield single-time expectation values, the Feynman-Vernon approach naturally provides stochastic Langevin equations. This explicit continuous noise representation is fundamentally better suited for monitoring the classical trajectory and extracting continuous metrological bounds via classical signal processing techniques.}

Building on the seminal developments of quantum Brownian motion \cite{caldeira1983path}, this path integral formulation allows us to directly extract the quantum-induced classical stochastic dynamics and more generally to investigate the decoherence of the mechanical oscillator caused by a bath of qubits \cite{stamp1994dissipation, stamp2006decoherence, unruh2000false, anglin1997deconstructing, zurek2003decoherence}. We find that a single qubit induces detectable effective deterministic and noise forces in the classical dynamics of the mesoscopic oscillator.

Moreover, these forces are state-dependent, notably transducing information from the qubit's state onto the motion of the classical oscillator. This is the spin-analogue effect of quantum-induced stochastic graviton noise in quantum gravity ~\cite{hu2008stochastic, parikh2021quantum}, and may have implications to metrology ~\cite{pistolesi2021proposal, lee2017topical,montenegro2019mechanical, wang2023coherent, burd2024experimental, santos2017optomechanical}, as well as tabletop quantum gravity experiments seeking to detect gravitational-induced entanglement \cite{spin_collab}.



\begin{figure}
    \centering
    \includegraphics[width=8.4 cm]{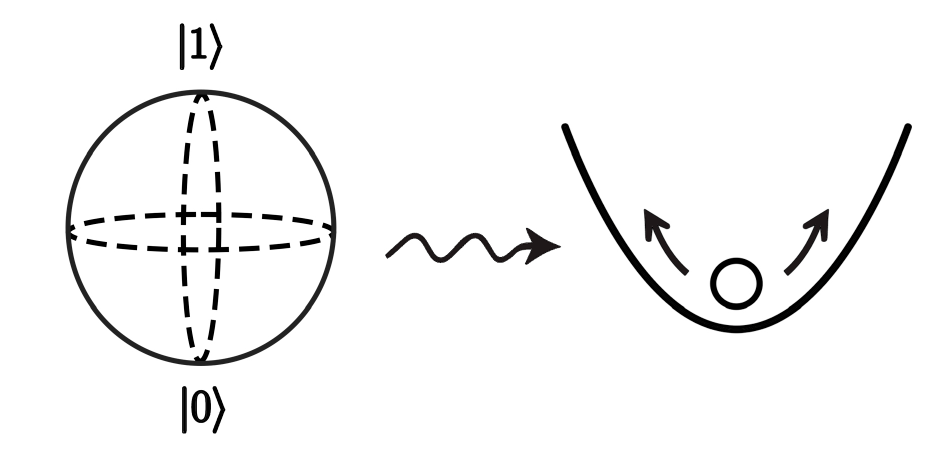}
    \caption{Conceptual representation of our quantum-classical system: a qubit interacting with a mesoscopic mechanical oscillator. }
    \label{fig:enter-label}
\end{figure}



This paper is organized as follows. Sec.~\ref{sec: hamiltonian} introduces the oscillator-qubit model used throughout this work. Then, in Sec.~\ref{sec: feynman-vernom} we provide details on the application of the Feynman-Vernon formalism to that model. Sec.~\ref{sec: quantum dynamics} derives the qubit-induced stochastic dynamics of the mechanical oscillator, highlighting how information on the qubit's state gets imprinted upon the classical motion, and computes the order-of-magnitude of the quantum-induced forces for different systems presented in the literature. In Sec.~\ref{sec:fisher}, we derive the Fisher Information Matrix for the qubit state through the continuous classical trajectory, quantifying the sensitivity of the oscillator to the initial quantum state. Finally, Sec.~\ref{sec: final remarks} discusses the broader implications of our findings to mesoscopic optomechanical systems and closes with a brief discussion of the results.

\section{Hamiltonian}\label{sec: hamiltonian}

We model the interaction between the qubit and the oscillator via the Jaynes-Cummings (JC) Hamiltonian~\cite{scully1997quantum}, which can be derived from a dipole interaction under the rotating wave approximation \cite{grynberg2010introduction}. Throughout, we will assume a single mode mechanical oscillator.

The total Hamiltonian is,
\begin{eqnarray}
    {H} = H_o + H_q + H_{\rm int},
\end{eqnarray}
where
\begin{eqnarray}
    {H_o} = {\hbar}\omega_o a^\dagger a ,~ {H_q} = \frac{1}{2}{\hbar}\omega_q \sigma_z \\
    H_{\rm int} = \frac{1}{2}{\hbar}\Omega\big(a\sigma_+ + a^\dagger\sigma_-\big), \label{eq: JC Hamiltonian}
\end{eqnarray}
denote the free and interaction terms. Here, $ \omega_{o}$, $ \omega_{q} $ and $\Omega$ denote the oscillator, qubit and the coupling frequencies, $ a $ ($ a^{\dagger}$) is the oscillator's annihilation (creation) operator, $\sigma_k$ represents the qubit Pauli matrices $(k = x,y,z)$ and $\sigma_\pm = \sigma_x \pm i \sigma_y$ are the qubit raising and lowering operators. We define the oscillator’s dimensionless quadrature operators as
\begin{equation}
    a = \frac{1}{\sqrt{2}}(q + i p), ~    a^\dagger = \frac{1}{\sqrt{2}}(q - i p),
\end{equation}
where $x = q_0q$ and $p_x =  p_0~p$ are the position and momentum operators, with $ q_{0} = \sqrt{\hbar/2m\omega_{o}}$ and $ p_{0} = \sqrt{m\hbar \omega_{o}/2}$ the zero-point motion position and momentum, with $q_0p_0 = \hbar/2$. We may rewrite the Hamiltonian in terms of the dimensionless quadrature operators,
\begin{eqnarray}
    {H_o} = \frac{1}{2}\hbar\omega_o\big(p^2 + q^2\big) ,~ {H_\text{int}} &=& \frac{\hbar\Omega}{2\sqrt{2}}\big(q\sigma_x - p\sigma_y\big) \ , \label{eq: new hamiltonians}
\end{eqnarray}
which provide a form naturally suited to describe the oscillator's classical phase-space dynamics.

\section{Feynman-Vernon formalism}\label{sec: feynman-vernom}

We now apply the Feynman-Vernon formalism \cite{feynman1963theory, paraguassu2024quantum}, with the goal of deriving the effective action of the oscillator induced by interacting with the qubit.



\subsection{Density matrix}

The starting point of the Feynman-Vernon influence functional method is a path integral representation for the system's density matrix \cite{feynman1965path}. We write,
\begin{equation}
    \rho(q,q',t) = \sum_{i=0,1} \langle{q,i}|U \rho_0 U^\dagger |q',i\rangle,    
\end{equation}
where $|q,i\rangle \equiv |q\rangle\otimes|i\rangle$  denotes an element of the basis formed by taking the tensor product between the oscillator's position eigenstates $\ket{q}$ and the qubit's canonical basis $\ket{i}$, which we take to be eigenstates of $ \sigma_{z}$, $U = \exp(-iHt/\hbar)$ is the unitary evolution of the system and $\rho_0$ is the initial density matrix, assumed to be a product state $\rho_0 = \rho_o \otimes \rho_q$. The assumption of an initially uncoupled qubit-oscillator system is justified provided each subsystem is prepared independently \cite{caldeira1983path}. We denote $q(t)$ and $p(t)$ the forward (``ket'') position and momentum variables, while $q'(t)$ and $p'(t)$ denote the backward (``bra'') position and momentum variables. 

Inserting a complete basis in the position and momentum variables~\cite{chaichian2018path}, the reduced density matrix can be written in path integral form,
\begin{eqnarray}
    \rho(q,q',t) &=& \sum_{i,j,k}\int dq_0 dq'_0\int \mathcal{D}q\mathcal{D}q'\mathcal{D}p \mathcal{D}p' \langle i| e^{i S(p,q)}|j\rangle \nonumber \\
    &&\times \langle q_0,j|\rho_0|q'_0,k\rangle\langle k|e^{-iS(p',q')}|i\rangle \nonumber
\end{eqnarray}
where $S(p,q) = \int_{0}^{t}dt' \big(p(t')\dot{q}(t') - \frac{1}{\hbar}H(p,q)\big) \equiv S$ and $  S' \equiv S(p',q')  $.
For clarity, we have written $\int\mathcal{D}q$ meaning $\int_{q_0}^{q}\mathcal{D}q$, to shorten the notation, analogously to the momentum paths.

Assuming a separable initial state $\rho_0 = \rho_o\otimes\rho_q$ and denoting the free oscillator action as $S_o = \int_{0}^{t}dt' \big(p(t')\dot{q}(t') - \frac{1}{\hbar}H_o(p,q)\big)$, we arrive at
\begin{eqnarray}
    \rho(q,q',t) &=& \int dq_0 dq'_0 \langle q_0 |\rho_o|q'_0\rangle \label{eq: density matrix}\\
    &&\times\int\mathcal{D}q\mathcal{D}q'\mathcal{D}p\mathcal{D}p' e^{i(S_o - S'_o)}\mathcal{F}[q,p,q',p'] \nonumber
\end{eqnarray}
where $\mathcal{F}[q,p,q',p']$ denotes the Feynman-Vernon influence functional given by,

\begin{eqnarray}\label{eq: pre influence}
    \mathcal{F}[q,p,q',p'] 
    &=& \sum_{i,j,k} \bra{i} e^{-\frac{i}{\hbar}\int dt\big(H_q + H_{\text{int}}(p,q)\big)}\ket{j} \\
    & \times& \bra{j}\rho_q\ket{k}\bra{k}e^{+\frac{i}{\hbar}\int dt\big(H_q + H_{\text{int}}(p',q')\big)}\ket{i}. \nonumber
\end{eqnarray}
For simplicity, we consider the initial qubit state to be pure, $\rho_q = |\Psi\rangle\langle\Psi|$. Note this assumption can be generalized to an arbitrary density matrix through a convex combination of pure-state influence functionals \cite{feynman1963theory}. The influence functional can be expressed in terms of a path-dependent evolution of the qubit operator, $U_{A}(X)$, where $X$ represents the oscillator's entire path variables $\{q,p\}$,
\begin{equation}\label{eq: influence functional}
    \mathcal{F}[X,X'] = \langle \Psi | U^{\dagger}_{A}(X')U_{A}(X)|\Psi\rangle.
\end{equation}
In the interaction picture, the evolution operator is given by~\cite{parikh2021signatures}
\begin{equation}\label{eq: atom time evolution operator}
    U_{A,I}(t_f,0;X) = \mathcal{T} \exp\left(-\frac{i}{\hbar}\int_{0}^{t_f}dt~H_{\text{int},I}(t)\right),
\end{equation}
where $\mathcal{T}$ denotes the time-ordering operator.

The interaction Hamiltonian is explicitly given by
\begin{equation}\label{eq: interaction Hamiltionian v1}
    H_{\text{int},I} = \frac{\hbar\Omega}{2\sqrt{2}} \Big( \sigma_{x,I}(t)f_x(t) - \sigma_{y,I}(t)f_y(t) \Big).
\end{equation}
where the Pauli operators in the interaction picture are
\begin{align}
    \sigma_{x,I}(t) &= \sigma_x \cos(\omega_q t) - \sigma_y \sin(\omega_q t), \\
    \sigma_{y,I}(t) &= \sigma_x \sin(\omega_q t) + \sigma_y \cos(\omega_q t),
\end{align}
and we have defined the path-dependent functions
\begin{align}
    f_x(t) &= q(t)\cos(\omega_q t) - p(t)\sin(\omega_q t), \label{eq: fx}\\
    f_y(t) &= q(t)\sin(\omega_q t) + p(t)\cos(\omega_q t). \label{eq: fy}
\end{align}
From the qubit's perspective, the functions $f_{i}(\tau)$, for $i=x,y$, are classical time-dependent coefficients (c-numbers), as they depend on oscillator's paths. Therefore, the interaction Hamiltonian in Eq. (\ref{eq: interaction Hamiltionian v1}) describes a qubit driven by a classical force, where $f_x(t)$ and $f_y(t)$ are the drive components. An explicit calculation of the evolution operator is detailed in Appendix~\ref{appendixA}.

\subsection{Influence Functional}

The influence functional eliminates all qubit degrees of freedom, leaving only the path integral over the oscillator's forward and backward variables $q$, $p$ and $q'$, $p'$. Introducing the dimensionless coupling 
\begin{eqnarray}
    g = \frac{\Omega}{2\sqrt{2}\,\omega_q}
\end{eqnarray}
and the rescaled time variable $\tau = \omega_q t$, the influence functional becomes(see Appendix \ref{appendixA} for details) 
\begin{equation}\label{eq: influence functional version 1}
    \mathcal{F}[X,X'] \approx \langle \Psi| e^{ig W_x\sigma_x }e^{ig  W_y\sigma_y}e^{ig^2W_z\sigma_z} |\Psi\rangle,
\end{equation}
where,
\begin{eqnarray}
    W_x &=& F'_x - F_x, \label{eq: Wx}\\
    W_y &=& F_y - F'_y, \label{eq: Wy}\\
    W_z &=& F'_z + 2F'_xF_y - F'_xF'_y -F_xF_y - F_z, \label{eq: Wz}
\end{eqnarray}
and the functionals $F_i$ are given by,
\begin{eqnarray}
    F_x &=& \int_{0}^{T}d\tau \, f_x(\tau), \\
    F_y &=& \int_{0}^{T}d\tau \, f_y(\tau), \\
    F_z &=& \int_{0}^{T}d\tau\int_{0}^{\tau}d\tau' \Big( f_x(\tau)f_y(\tau') - f_x(\tau')f_y(\tau) \Big),
\end{eqnarray}
with $f_{i}(\tau)$ for $i=x,y$ defined in Eqs. \eqref{eq: fx} and \eqref{eq: fy}. 
{Despite the apparent complexity of the influence functional, the final form is achieved through successive use of the BCH formula, alongside with computations of commutators of the Pauli matrices.} 
Note that the resulting form of Eq.~\eqref{eq: influence functional version 1} differs from the influence functional obtained from the linear coupling between two oscillators \cite{paraguassu2024quantum}, which is given by $\mathcal{F}[q,q']= \bra{\Psi} e^{a^\dagger W} e^{a W^*}\ket{\psi}$, and is subject to the algebra of creation and annihilation operators. In the qubit case, instead of ladder operators, the functionals $W_i$ multiply each of the Pauli matrices, which satisfy the angular momentum algebra. This will significantly impact the effective dynamics of the oscillator.

We parametrize an arbitrary pure qubit state as,
\begin{equation}\label{eq: atom inital state}
    |\Psi\rangle = \sqrt{1-p}|0\rangle + e^{i\varphi}\sqrt{p}|1\rangle,
\end{equation}
where $p \in [0,1]$ and $\varphi \in [0,2\pi)$. 
To leading order in the dimensionless coupling $g$ we find,

\begin{equation}\label{eq: influence functional final version}
    \mathcal{F}[X,X'] \approx e^{i \Phi^{\text{fl}}}e^{i \Phi^{\text{forces}}},
\end{equation}
where we define the \textit{influence phases}, 
\begin{widetext}
    \begin{eqnarray}
        i \Phi^{\text{fl}} 
        &=& - \frac{g^2}{2}\Big(W^2_x + W^2_y - 2p(1-p)\big( W^2_x(1+\cos(2\varphi)) + W^2_y(1-\cos(2\varphi)) + 2 W_x W_y \sin(2\varphi) \big)  \Big), \label{eq: fluctuation influence phase}\\
        i \Phi^{\text{forces}} &=& 2ig\sqrt{p(1-p)}\Big(W_x\cos\varphi + W_y \sin\varphi \Big) + i g^2(1-2p)\big(W_z - W_x W_y\big), \label{eq: dissipation influence phase}
    \end{eqnarray}
\end{widetext}
The phase $i\Phi^{\text{fl}}$ contains fluctuations terms, while $i\Phi^{\text{forces}}$  describes dissipation and deterministic forces caused by the qubit upon the oscillator. 
Note that the influence functional can be trivially generalized to the case of $n$ non-interacting qubits, becoming $\mathcal{F}_n[X,X'] \approx e^{i n\Phi^{\text{fl}}}e^{i n\Phi^{\text{forces}}}$ \cite{feynman1963theory}. We now proceed to analyze each of these phases individually. 

In the following, we will give a sketch of the calculation. Since we have the influence functional written in terms of Pauli matrices, we simply apply the resulting operators and explicitly trace out the qubit, as usually done in quantum information theory \cite{nielsen2010quantum}. It is worth noticing that this is a lengthy but rather straightforward calculation relying only on the algebra of the Pauli matrices.
We refer to Appendix \ref{appendixB} and the Supplemental Material for further details on the influence phases.

\subsection{Fluctuation and Decoherence}

The fluctuation phase $ i\Phi^{\text{fl}} $ will contribute noise terms to the oscillator's dynamics. Its contribution is determined by $W^2_x$, $W^2_y$ and $W_xW_y$. The fluctuation phase can be put in the form,
\begin{equation}\label{eq: fluctuation influence phase with vectors}
    i \Phi^{\text{fl}} = -\frac{g^2}{2}\int_{0}^{T} d\tau d\tau' J^{T}(\tau)\mathcal{M}_{\text{fl}}(\tau,\tau')J(\tau')
\end{equation}
where $J(\tau)$ is a vector given by,
\begin{equation}\label{eq: vector J}
    J(\tau) = \left( \begin{array}{c}
    Q(\tau) \\
    P(\tau)
    \end{array} \right),
\end{equation}
with,
\begin{equation}
    Q(\tau) = q(\tau) - q'(\tau), ~
    P(\tau) = p(\tau) - p'(\tau).
\end{equation}
and $\mathcal{M}_{\text{fl}}(\tau,\tau')$ denotes the two-time noise kernel matrix.

Defining the matrices $M_{ij}(\tau,\tau')$, where
\begin{eqnarray}
    &&M_{xx}(\tau,\tau') = \left( \begin{array}{rr}
        \cos\tau \cos \tau' &  -\cos\tau \sin \tau' \\
        -\sin\tau \cos \tau'  & \sin\tau \sin \tau' 
    \end{array} \right), \\
    &&M_{yy}(\tau,\tau') = \left( \begin{array}{rr}
        \sin\tau \sin \tau' &  \sin\tau \cos \tau' \\
        \cos\tau \sin \tau'  & \cos\tau \cos \tau' 
    \end{array} \right),\\
    &&M_{xy}(\tau,\tau') = \left( \begin{array}{rr}
        -\cos\tau \sin \tau' &  -\frac{1}{2}\cos(\tau +\tau') \\
        -\frac{1}{2}\cos(\tau +\tau')  & \sin\tau \cos \tau' 
    \end{array} \right), ~~~~~~
\end{eqnarray}
the full noise kernel matrix is then given by, \begin{eqnarray}\label{eq: matrix M}
        \mathcal{M}_{\text{fl}}(\tau,\tau') &=& M_{xx} + M_{yy} - 2p(1-p)\Big[ M_{xx}(1+\cos(2\varphi)) \nonumber \\
        &+& M_{yy}(1-\cos(2\varphi)) + 2 M_{xy} \sin(2\varphi) \Big].
\end{eqnarray}


As noted in Refs. \cite{weiss2012quantum,calzetta1988nonequilibrium}, Eq.~\eqref{eq: fluctuation influence phase with vectors} represents decoherence of the oscillator induced by the interaction with the qubit. The effect of $i \Phi^{\text{fl}} $ is to suppress off-diagonal elements in the density matrix (for which $q\neq q', p\neq p'$); this term is sometimes also called the decoherence functional \cite{dowker1992quantum}. Note in our case the appearance of additional momentum terms. In the Wigner representation, these terms manifest as a diffusion process, providing an alternative perspective on the quantum-to-classical transition dynamics \cite{calzetta1988nonequilibrium, zurek1991decoherence}. 

For the classical regime of interest, the fluctuation term can be interpreted as a noise contribution. To see that, we perform a generalization of the so-called Feynman-Vernon-Stratonovich trick \cite{feynman1963theory}, which consists in writing Eq. \eqref{eq: fluctuation influence phase with vectors} in terms of a path integral over an auxiliary ``noise vector'' $\Lambda(\tau)$,
\begin{eqnarray}
    &&e^{-\frac{g^2}{2}\int_{0}^{T} d\tau d\tau' J^{T}(\tau)\mathcal{M}(\tau,\tau')J(\tau')} \label{eq: Feynman trick}\\
    &&=\int \mathcal{D}\Lambda e^{-\frac{1}{2}\int_{0}^{T} d\tau d\tau' \Lambda^{T}(\tau)\mathcal{M}^{-1}(\tau,\tau')\Lambda(\tau') + i \int_{0}^{T} d\tau \Lambda^{T}(\tau)J(\tau)} \nonumber
\end{eqnarray}
where $\int_{0}^{T} d\tau'' \mathcal{M}(\tau,\tau'')\mathcal{M}^{-1}(\tau'',\tau') = \delta(\tau-\tau')$. This operation decouples the forward and backward variables (defined in terms of the vectors $J$ and $J'$), at the expense of introducing $\Lambda$. Note that while $i \Phi^{\text{fl}} $ appears at second order, the Feynman trick yields a term linear in $ g $. 

From Eq. \eqref{eq: Feynman trick} we may read out the joint probability density functional of the noise vector $\Lambda(\tau)$,
\begin{equation}
    P[\Lambda(\tau)] = e^{-\frac{1}{2}\int_{0}^{T}d\tau d\tau'\Lambda^{T}(\tau) \mathcal{M}^{-1}_{\text{fl}}(\tau,\tau')\Lambda(\tau')}
\end{equation}
where we define the noise terms,
\begin{equation}\label{eq: noise vector}
    \Lambda(\tau) \equiv g\left( \begin{array}{c}
    \lambda_q(\tau) \\
    \lambda_p(\tau)
    \end{array} \right).
\end{equation}
with mean and correlation function given by
\begin{equation}
    \langle \Lambda(\tau) \rangle = 0, ~\langle \Lambda_i(\tau)\Lambda_j(\tau') \rangle = \mathcal{M}_{ij}(\tau,\tau') \ ,
\end{equation}
or equivalently,
\begin{equation}\label{eq: lambda averages}
    \langle \lambda_q(\tau) \rangle =\langle \lambda_p(\tau) \rangle = 0, ~\langle \lambda_i(\tau)\lambda_j(\tau') \rangle = \mathcal{M}_{ij}(\tau,\tau'),
\end{equation}
where $\mathcal{M}_{ij}(\tau,\tau')$ are the elements of the matrix $\mathcal{M}_{\text{fl}}(\tau,\tau')$ defined in Eq. \eqref{eq: matrix M}. We see that the stochastic variables $\lambda_q$ and $\lambda_p$ are correlated.

Finally, denoting the stochastic average over $P[\Lambda(\tau)]$ by $\langle \dots \rangle$, Eq. \eqref{eq: Feynman trick} can be interpreted as the average of
\begin{equation}
    \langle e^{i \int_{0}^{T} d\tau \Lambda^{T}(\tau)J(\tau)} \rangle.
\end{equation}

\subsection{Dissipation and Forces}

The phase $i\Phi^{\text{force}}$ contains terms associated with both dissipation and deterministic forces. The dissipative component of Eq. \eqref{eq: dissipation influence phase} is quadratic in the coupling constant $g$, and is given by the term,
\begin{equation}
    i g^2(1-2p)\big(W_z - W_x W_y\big) \ ,
\end{equation}
which arises from the interaction between the qubit and the oscillator's momentum and represents a form of generalized dissipation~\cite{parikh2021signatures, calzetta1988nonequilibrium}. Unlike the fluctuation phase $i\Phi^{\text{fl}}$, this term is not bilinear in the oscillator's coordinates, which prevents us from using the `Feynman trick' discussed above. In this way, while the fluctuation term contributes in the effective dynamics at first order in $g$, the dissipative term persists as a second order effect in the semiclassical equation of motion. The validity of this approximation can be readily seen in Table \ref{tab:exp_params}, due the value of $g$ for different experimental parameters. It is worth noticing that the full calculation would lead to other higher order effects, such as a frequency shift of the oscillator.  These terms are subleading for the cases analyzed since $g\lesssim0.1$. Hence, in the weak coupling regime, these contributions will be neglected from now on~\cite{paraguassu2024quantum}. 

Focusing on the first-order contribution to $i\Phi^{\text{force}}$, we write
\begin{equation}
    i\Phi^{\text{force}} = -i \int_{0}^{T}d\tau \,\mathcal{F}^{T}_{\text{force}}(\tau)J(\tau),
\end{equation}
where the deterministic force vector $\mathcal{F}_{\text{force}}(\tau)$ is
\begin{equation}\label{eq: M dissipation vector}
    \mathcal{F}_{\text{force}}(\tau) = {2} g\sqrt{p(1-p)}\left( \begin{array}{rr}
        \cos(\tau+\varphi) \\
        -\sin(\tau+\varphi)
    \end{array}\right).
\end{equation}
\\
Observe that this force is proportional to $\sqrt{p(1-p)}$, meaning it exists only when the qubit is in a superposition of the $\sigma_{z}$ eigenstates. 


\section{Quantum-induced stochastic dynamics}\label{sec: quantum dynamics}

\subsection{Super-propagator}

Combining the above results, we can write the density matrix in its final form,
\begin{eqnarray}
    \rho(q,q',t) &=& \int dq_0 dq'_0 \rho_0(q_0,q_0')\mathcal{J}(X_f,X'_f|X_0,X'_0) ~~~~
\end{eqnarray}
where the \textit{super-propagator} is defined as, 
\begin{widetext}
\begin{eqnarray}
    \mathcal{J}(X_f,X'_f|X_0,X'_0) &=& \int_{X_0,X'_0}^{X_f,X'_f}\mathcal{D}X\mathcal{D}X'\int \mathcal{D}\Lambda~ \exp\Big(-\frac{1}{2}\int_{0}^{T}d\tau d\tau'\Lambda^{T}(\tau) \mathcal{M}^{-1}_{\text{fl}}(\tau,\tau')\Lambda(\tau')\Big) \label{eq: propagator}\\
    && \times \exp\Bigg({i} \int_{0}^{T}d\tau \Big( p(\tau)\dot{q}(\tau) -p'(\tau)\dot{q}'(\tau) - \frac{1}{ \hbar\omega_q} \big(H_o(\tau) - H'_o(\tau) \big)+\big( \Lambda(\tau) -\mathcal{F}_{\text{force}}(\tau)\big)^{T}J(\tau)\Big) \Bigg). \nonumber
    \end{eqnarray}

\end{widetext}

\noindent Here, the first line represents the stochastic density kernel, while the terms in the second line represent the oscillator's forward and backward free actions written in Hamiltonian form, as well as the stochastic and deterministic forces caused by the qubit. 




\begin{figure}[ht!]
    \centering
    \includegraphics[width=8.6 cm]{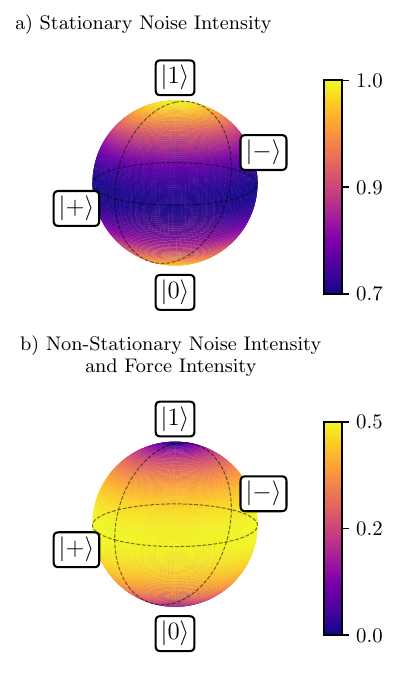}
    \caption{State dependence of the deterministic and stochastic forces induced by the qubit on the mechanical oscillator, as visualized in the Bloch sphere. State-dependent factor for the (a) stationary stochastic force, $\eta_{st} $ (see Eq. \eqref{eq: force intensity}), achieving maximal value at the poles (the states $|0\rangle$ and excited $|1\rangle$ states) and for the (b) non-stationary stochastic and deterministic forces, $\eta_f $ (see Eq. \eqref{eq: force intensity2}), which vanish at the poles and are maximal for superposition states along the equator. Non-stationary stochastic and deterministic forces are direct signatures of quantum superposition.}
    \label{fig:bloch sphere}
    \end{figure}

\subsection{Equations of motion}

The quantum-induced classical stochastic dynamics of the oscillator is obtained by
extremising the argument in the super-propagator defined in Eq. \eqref{eq: propagator} with respect $q(\tau)$, $q'(\tau)$, $p(\tau)$ and $p'(\tau)$, which leads to effective Hamilton equations for the forwards and backwards paths \cite{paraguassu2024quantum, parikh2021signatures}. Neglecting dissipation, the forward and backward variables obey the same uncoupled equations of motion given by,
\begin{eqnarray}
    \dot{q} - r p + g\left( 2\sqrt{p(1-p)} \sin(\tau+\varphi) + \lambda_p \right) &=& 0, \label{eq:q_final}\\
    \dot{p} + r q + g \left(2\sqrt{p(1-p)} \cos(\tau + \varphi) - \lambda_q\right) &=& 0, \label{eq:p_final}
\end{eqnarray}
where $r = \omega_o/\omega_q$. From these we can obtain the Eq. of motion for $q(\tau)$,
\begin{equation}\label{eq_motion}
    \ddot{q} + r^2 q = -g\Big(2\sqrt{p(1-p)}\left( 1+r\right)\cos(\tau+\varphi) + \dot{\lambda_p} - r\lambda_q \Big). 
\end{equation}
The mean value and the correlation functions of $\lambda_{p,q}$ given by Eq. \eqref{eq: lambda averages} can be written explicitly,
\begin{eqnarray}
    \langle \lambda_q(\tau)\lambda_q(\tau') \rangle &=&  \big(\big(1- p\big)^2+p^2\big) \cos(\tau-\tau')   \label{eq: noise correlation}\\
    &&-2p(1-p) \cos(\tau+\tau'+2 \varphi)\nonumber \\
    &&~~~~~~~~~~~~~~~~ \nonumber\\   
    \langle \lambda_p(\tau)\lambda_p(\tau') \rangle &=& \big(\big(1- p\big)^2+p^2\big) \cos(\tau-\tau')   \\
    &&+2p(1-p)\cos(\tau+\tau'+2 \varphi)\nonumber \\
    &&~~~~~~~~~~~~~~~~ \nonumber\\
    \langle \lambda_q(\tau)\lambda_p(\tau') \rangle
    &=& \sin(\tau-\tau')\\
    && +2p(1-p)\sin(\tau+\tau'+2\varphi). \nonumber
\end{eqnarray}

\noindent Note that besides correlations, we have non-stationary noise components which remarkably only appear when the initial qubit state is in a superposition of the $ \vert 0 \rangle$, $ \vert 1 \rangle$ states, since for $p=0$ or $p=1$ these non-stationary contributions vanish. Observe that non-stationary noise also appears in the case of two oscillators, where one of the oscillators is initially in a squeezed state~\cite{parikh2021signatures,paraguassu2024quantum, paraguassu2022probabilities}.
Moreover, the oscillator experiences a driving force that critically depends on the qubit's state in the Bloch sphere. Remarkably, this force reaches its maximum value when $p=1/2$, while vanishing for $p=0$ and $p=1$; compare this to the non-stationary noise components.

Finally, we observe that the effect of the deterministic force can be used to gain information on the qubit's initial state, since this force depends on the parameters $ p $ and $ \varphi $, and affects the mean value of the oscillator's position, 
\begin{eqnarray}
    &&\langle q(\tau)\rangle = -\frac{2g}{1-r}\sqrt{p(1-p)} \label{eq: average of q} \\
    &&\times\left[\cos\varphi\big( \cos(r \tau)-\cos\tau \big) - \frac{1}{r}\sin\varphi\big(\sin(r \tau)-\sin\tau \big)\right], \nonumber 
\end{eqnarray}
for $q(0) = \dot{q}(0) = 0$.

This result allows one to partially reconstruct the qubit's state by monitoring the oscillator's classical position. Higher-order moments of the oscillator's dynamics will also be influenced by the qubit's initial state, opening the way to quantum state reconstruction via classical noise spectroscopy.

\subsection{Application to spin-optomechanical systems}
We now apply the results described thus far to spin-optomechanical systems described by the JC Hamiltonian, notably trapped ions \cite{meekhof1996generation}, levitated nanodiamonds \cite{yin2013large} and piezoelectric mechanical oscillators coupled to superconducting qubits \cite{bild2023schrodinger}.

Define the characteristic force, 
\begin{eqnarray}\label{eq: characteristic force}
    \mathbf{f}_{0} \equiv \frac{\hbar \Omega}{4\sqrt{2}q_{0}} \ ,
\end{eqnarray}
which governs the order-of-magnitude of forces induced by the interaction with the qubit. Restoring dimensions to Eq.~\eqref{eq_motion}, we find 
\begin{equation}\label{eq: dimensional oscillator}
    m\frac{d^2x}{dt^2} + m\omega^2_o x  = -f(t) + \xi_q(t)-\xi_p(t),
\end{equation}
where $f(t)$ corresponds to the deterministic force, while $ \xi_{q}, \xi_{p}$ describe the stochastic quantum-induced forces. These are given by,
\begin{eqnarray}
    f(t) &=& 2\mathbf{f}_{0} \left(\frac{1+r}{r}\right) \eta_{f}(p) \cos(\omega_{q}t + \varphi) \label{eq:deterministic_force}\\
    \xi_{q} &=& \mathbf{f}_{0} \lambda_{q}(t) \\
    \xi_{p} &=& \frac{\mathbf{f}_{0}}{\omega_{o}} \frac{d}{dt} \lambda_{p}(t)
\end{eqnarray}
where we have defined the state-dependent intensity factor 
\begin{eqnarray}
    {\eta_{f}(p) \equiv \sqrt{p(1-p)}}\label{eq: force intensity}
\end{eqnarray}
which governs the magnitude of the deterministic force,
and the stochastic forces have correlators given by, 
\begin{eqnarray}
    \langle \xi_q(t) \xi_q(t') \rangle &=& \mathbf{f}_{0}^2\langle \lambda_q(\tau)\lambda_q(\tau') \rangle, \label{eq: dimensional noise correlation}\\   
    \langle \xi_p(t) \xi_p(t') \rangle &=& \frac{\mathbf{f}_{0}^{2}}{r^{2}}\langle \lambda_p(\tau')\lambda_p(\tau) \rangle, \\
    \langle \xi_q(t)\xi_p(t') \rangle
    &=& -\frac{\mathbf{f}_{0}^{2}}{r}\Big[ \cos(\tau-\tau')\label{eq: dimensional noise correlation2}\\
    && ~~~~~~~ - \eta^2_f \cos(\tau+\varphi)\cos(\tau'+\varphi)\Big]. \nonumber
\end{eqnarray}
Looking at Eq.~\eqref{eq: noise correlation}, we find the magnitude of the non-stationary component of the stochastic force is also governed by the state-dependent factor $ \eta_{f}(p) $, while the stationary component has the accompanying magnitude intensity factor
\begin{equation}
\eta_{st}(p) \equiv \sqrt{(1-p)^2 + p^2} \ .
\label{eq: force intensity2}
\end{equation}

We can visualize the state-dependence of both the deterministic and stochastic forces by plotting $ \eta_{st} $ and $ \eta_{f} $ in the Bloch sphere. To do that, we parametrize the state probability $ p = \sin^2 ( \theta/2) $ in terms of the latitude angle $ \theta$. Note that the azimuthal angle $ \varphi $ only affects the phase of the deterministic and non-stationary forces [see Eqs. \eqref{eq_motion} and ~\eqref{eq: noise correlation}]. The state-dependent factors are depicted in Fig.~\ref{fig:bloch sphere}. 

\begin{table}[t!]
\caption{\label{tab:exp_params}Comparison of experimental parameters and associated quantum-induced spin-optomechanical forces. Note that since the qubit and oscillator frequencies coincide in the case of Ref. \cite{bild2023schrodinger} the deterministic force vanishes.}
\begin{ruledtabular}
\begin{tabular}{cccc}
Parameter &  Ion \textbf{\cite{meekhof1996generation}} & Nanodiamond \textbf{\cite{yin2013large}} & Piezo \textbf{\cite{bild2023schrodinger}} \\
\hline 
$m$ (kg)              & $\;\;\;1.5 \times 10^{-26}$ & $\;\;\;5.5 \times 10^{-17}$ &  $\;\;\;1.6 \times 10^{-8}$ \\
$\Omega/2\pi$ (kHz)   & $5.0 \times 10^{2}$ & $5.2 \times 10^{1}$ &  $1.6 \times 10^{3}$ \\
$\omega_o/2\pi$ (MHz) & $1.1 \times 10^{1}$ &$\;\;\,5.0 \times 10^{-1}$ &  $1.2 \times 10^{1}$ \\
$\omega_q/2\pi$ (MHz) &  $1.2 \times 10^{3}$ &$\;\;\,2.5 \times 10^{-1}$ & $1.2 \times 10^{1}$ \\
$g$                   & $1.5 \times 10^{-4}$ & $7.4 \times 10^{-2}$ & $4.7 \times 10^{-2}$ \\
\hline
$f_0$ (N)           & $1.8 \times 10^{-18}$ &$3.3 \times 10^{-17}$ &  $ 1.1 \times 10^{-10} $ \\
$\xi_{q,0}$ (N)       & $8.2 \times 10^{-21}$ &$1.1 \times 10^{-17}$ &  $2.8 \times 10^{-11}$ \\
$\xi_{p,0}$ (N)       & $9.0 \times 10^{-19}$ &$5.5 \times 10^{-18}$ &  $2.8 \times 10^{-11}$ \\
\end{tabular}
\end{ruledtabular}
\end{table}

Finally, we can estimate the order-of-magnitude of the quantum-induced forces for each of the optomechanical setups discussed in Refs. \cite{meekhof1996generation, yin2013large, bild2023schrodinger}. To do so, we define the following quantities carrying the magnitude of the forces,
\begin{equation}\label{eq: force magnitude}
    f_0 \equiv 2\mathbf{f}_{0}\left(\frac{\omega_q}{\omega_o}+1\right), ~\xi_{q,0} \equiv \mathbf{f}_{0}, ~\xi_{p,0} \equiv  {\mathbf{f}_{0}}\frac{\omega_q}{\omega_o} \ .
\end{equation}
Table \ref{tab:exp_params} shows the relevant mass, frequency and JC coupling parameters for each of the oscillators and qubits in Refs.~\cite{meekhof1996generation, yin2013large, bild2023schrodinger}, and the resulting magnitudes of the quantum-induced deterministic and stochastic forces. 

Crucially, this entire force range is detectable by various state-of-the-art experimental platforms, notably levitated systems and nanomechanical resonators~\cite{ranjit2016zeptonewton, monteiro2020force, weiss2012quantum, ranjit2015attonewton, gavartin2012hybrid, yan2023force}. We highlight these quantum-induced forces might be of special relevance to near-future spin-mechanical tabletop quantum gravity experiments with levitated nanodiamonds \cite{bose2017spin, spin_collab}. The explicit dependence of both the deterministic and stochastic forces on the initial state parameters $(\theta, \phi)$ indicates that the classical trajectory acts as a continuous measurement record. In the following section, we formalize this intuition by calculating the Fisher Information Matrix (FIM) to establish the fundamental precision limits for quantum state reconstruction.

\section{Estimating the qubit state through Fisher Information}\label{sec:fisher}

\begin{figure*}[htpb]
    \centering
    \includegraphics[width=\textwidth]{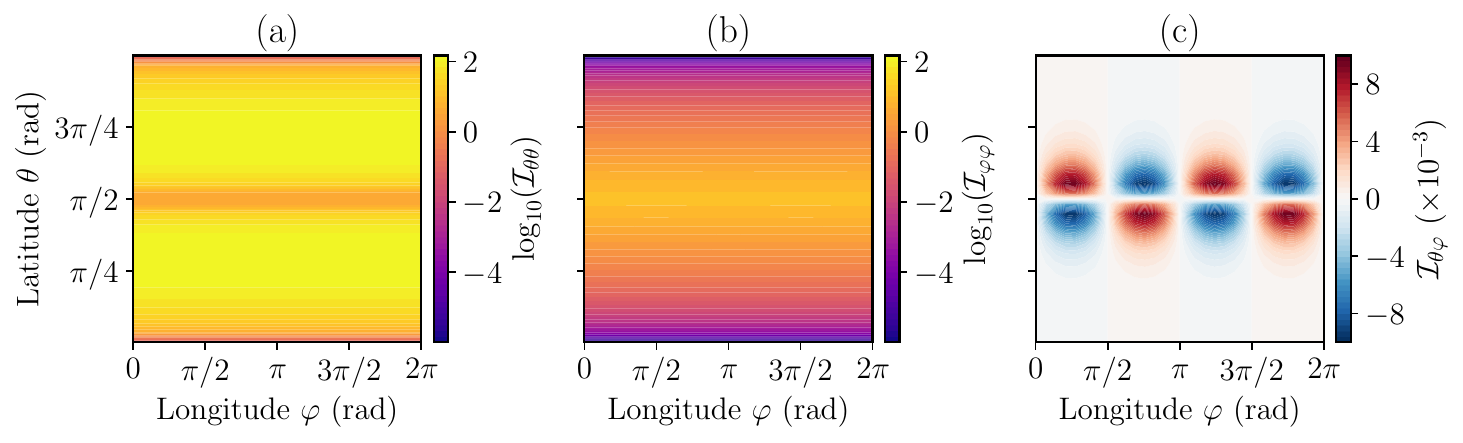}
    \caption{Fisher Information Matrix components for the estimation of the initial qubit state $(\theta, \phi)$ through the classical trajectory of the mechanical oscillator. We set $r = 0.1$, $g=0.1$, and $\Gamma = 10^{-3}$. (a) The information concerning the population parameter, $\mathcal{I}_{\theta\theta}$, which peaks at intermediate latitudes but drops at the equator. (b) The information concerning the phase, $\mathcal{I}_{\varphi\varphi}$, which is maximized for maximal superposition states ($\theta = \pi/2$). (c) The cross-information $\mathcal{I}_{\theta\varphi}$, illustrating regions where the estimation of $\theta$ and $\varphi$ are correlated. The zero-crossings (white regions) indicate states where the parameters can be estimated independently. For the numerical evaluation of the discrete Green's functions and covariance matrices, we consider a physical interaction time $t = 1$ ms and a qubit frequency $\omega_q/2\pi = 0.25$ MHz (consistent with the nanodiamond parameters in Table \ref{tab:exp_params}), yielding a total dimensionless time $T = \omega_q t \approx 1.57 \times 10^3$ discretized over $N = 1000$ steps.}
    \label{fig:fisher_maps}
\end{figure*}

To see the visibility of the qubit state through the stochastic classical oscillator, we need to compute the Fisher information for the qubit state depending on $p$ and \textcolor{blue}{$\varphi$}. The Fisher Information Matrix (FIM) for these parameters will show how sensitive the mechanical oscillator is to changes in the qubit's state, highlighting the possibility of using it as tool to partially reconstruct the quantum state.

First, we rewrite the stochastic dynamics{, given by Eq.\eqref{eq: dimensional oscillator},} in terms of a single effective noise and consider $p = \sin^2(\theta/2)$ {for $\theta \in [0,\pi]$} to simplify the analysis. The {state-dependent and magnitude intensity factors, given by Eqs.\eqref{eq: force intensity} and \eqref{eq: force intensity2} respectively, can now be written as}
\begin{align}
    \eta_f(\theta) &\equiv \frac{1}{2}\sin\theta, \label{eq: force intensity_theta}\\
    \eta_{st}(\theta)&\equiv \frac{1}{2}\sqrt{\cos(2\theta)+3} , \label{eq: force intensity2_theta}
\end{align}
{for $\theta \in [0,\pi]$. These quantities allows us to redefine the deterministic force given by Eq.\eqref{eq:deterministic_force}, as well as the stochastic correlations given by Eqs.\eqref{eq: dimensional noise correlation}-\eqref{eq: dimensional noise correlation2}. To make it clear, let us rewrite the dimensionless quantities here. Now, the deterministic force will be called by $f(t;\theta,\varphi)$,}
\begin{equation}
    f(t;\theta,\varphi) = 2\mathbf{f}_0 \left( \frac{1-r}{r} \right)\eta_f\cos(\tau+\varphi), 
\end{equation}
{and define a zero-mean noise given by $\eta \equiv \xi_q - \xi_p$ whose correlation is a combination between Eqs.\eqref{eq: dimensional noise correlation}-\eqref{eq: dimensional noise correlation2},} 
\begin{align}
    \langle \eta(t) \eta(t') \rangle = \frac{\mathbf{f}^2_0}{r^2}&\Big\{ \left[2r+(1-r+r^2)\eta_f^2 \right]\cos(\tau-\tau') \label{noise2a}  \\
    &+\left[2(1-r^2)\eta^2_{st}-r \eta^2_f \right]\cos(\tau+\tau'+2\varphi)\Big\}, \nonumber
\end{align}
{remembering that $\tau = \omega_q t$.}

In realistic experimental scenarios, a ``perfect'' qubit with an exact transition frequency $\omega_q$ is an idealization, as coupling to the environment and intrinsic physical imperfections introduce frequency fluctuations. Phenomenologically, we account for these decoherence mechanisms by broadening the transition frequency into a Lorentzian distribution centered at $\omega_q$, which inherently leads to an exponential decay of coherence \cite{yuge2011measurement, cywinski2008enhance, breuer2002theory}. Crucially, this physical broadening is also a mathematical necessity for metrology. An idealized pure cosine noise kernel, as in Eqs.\eqref{noise2a}, is strictly singular and its inverse is not well-defined. Integrating over the Lorentzian distribution regularizes the noise kernel by introducing a finite correlation time driven by the decay rate $\Gamma$. The regularized two-time correlation function and deterministic force are thus given by
\begin{align}
    f(t;\theta,\varphi) &= 2\mathbf{f}_0 \left( \frac{1-r}{r} \right)\eta_f e^{-\frac{\Gamma}{2}\tau}\cos(\tau+\varphi), \\
    \langle \eta(t) \eta(t') \rangle &= \frac{\mathbf{f}^2_0}{r^2}\Big\{ \left[2r+(1-r+r^2)\eta_f^2 \right]e^{-\frac{\Gamma}{2}|\tau-\tau'|\tau}\cos(\tau-\tau') \nonumber\\
    &+\left[2(1-r^2)\eta^2_{st}-r \eta^2_f \right]e^{-\frac{\Gamma}{2}(\tau+\tau')}\cos(\tau+\tau'+2\varphi)\Big\}, \nonumber
\end{align}
where we detail the calculation steps in Appendix \ref{appendixD}. 

Assuming the oscillator starts at rest, its dynamics can be formally solved using the retarded Green's function $G(\tau,\varsigma) = r^{-1}\sin[r(\tau-\varsigma)]\Theta(\tau-\varsigma)$, where $\Theta(\cdot)$ is the Heaviside step function. The mean trajectory $\mu(t) \equiv \langle q(t) \rangle$ and the two-time covariance $C_q(\tau,\tau') \equiv \langle q(\tau)q(\tau') \rangle - \mu(\tau)\mu(\tau')$ are given by convolutions with the deterministic force and the noise kernel, respectively
\begin{equation}
    \mu(\tau) = \int_0^\tau G(\tau,\varsigma) f(\varsigma; \theta, \varphi) d\varsigma,
\end{equation}
\begin{equation}
    C_q(\tau,\tau') = \int_0^\tau \int_0^{\tau'} G(\tau,\varsigma) G(\tau',\varsigma') \langle\eta(\varsigma)\eta(\varsigma')\rangle d\varsigma\, d\varsigma'.
\end{equation}

To properly perform quantum state tomography of the qubit parameters $\boldsymbol{\theta} = (\theta, \varphi)$, we calculate the FIM. Since the underlying process is Gaussian, the Slepian-Bangs formula can be employed \cite{kay1993fundamentals}. For numerical stability and practical evaluation, we discretize the time domain into $N$ steps of size $\Delta \tau$, such that $\tau_k = k \Delta \tau$. The continuous integral operators are then mapped into matrix multiplications.

Defining the lower-triangular Green's matrix as $[\mathbf{G}]_{kl} = G(\tau_k, \tau_l) \Delta \tau$, the discretized mean vector and covariance matrix are simply evaluated as
\begin{equation}
    \boldsymbol{\mu} = \mathbf{G} \mathbf{f}, \quad \text{and} \quad \mathbf{C}_q = \mathbf{G} \mathbf{C}_\eta \mathbf{G}^T,
\end{equation}
where $\mathbf{f}$ is the discretized force vector and $[\mathbf{C}_\eta]_{kl} = \langle\eta(\tau_k)\eta(\tau_l)\rangle$. Consequently, the parameter derivatives required for the Slepian-Bangs formula are straightforwardly obtained via linear propagation, $\partial_i \boldsymbol{\mu} = \mathbf{G} (\partial_i \mathbf{f})$ and $\partial_i \mathbf{C}_q = \mathbf{G} (\partial_i \mathbf{C}_\eta) \mathbf{G}^T$.

Finally, the discrete FIM elements are computed as \cite{kay1993fundamentals}
\begin{equation}
    \mathcal{I}_{ij} = (\partial_i \boldsymbol{\mu})^T \mathbf{C}_q^{-1} (\partial_j \boldsymbol{\mu}) + \frac{1}{2} \text{Tr}\left( \mathbf{C}_q^{-1} (\partial_i \mathbf{C}_q) \mathbf{C}_q^{-1} (\partial_j \mathbf{C}_q) \right),
\end{equation}
where $\partial_i \equiv \partial / \partial \theta_i$. 

To quantify the sensitivity of the classical oscillator to the initial state of the qubit, we numerically evaluate the elements of the FIM. In Fig. \ref{fig:fisher_maps}, we plot the FIM components $\mathcal{I}_{\theta\theta}$, $\mathcal{I}_{\varphi\varphi}$, and the cross-information $\mathcal{I}_{\theta\varphi}$ as functions of the initial qubit state $(\theta, \varphi)$ in the Bloch sphere. Following the physical parameters for levitated nanodiamonds outlined in Table \ref{tab:exp_params}, we set the frequency ratio to $r=0.1$. To model a realistic scale decoherence regime, we assume a dephasing time of $T_2 \approx 0.6$ ms, which corresponds to a physical decoherence rate of approximately $250$ Hz. This value aligns closely with state-of-the-art coherence times reported for high-purity nanodiamond NV centers~\cite{knowles2014observing} and modern superconducting transmon qubits~\cite{place2021new, somoroff2023millisecond}. Rescaling this rate by the chosen qubit frequency ($\omega_q/2\pi = 0.25$ MHz), we obtain the dimensionless decay parameter $\Gamma = 10^{-3}$ used in our simulations.

The estimation precision exhibits a strong state-dependence, as shown in Fig. \ref{fig:fisher_maps}. The Fisher information for the population angle, $\mathcal{I}_{\theta\theta}$ (Fig. \ref{fig:fisher_maps}(a)), peaks at intermediate latitudes (e.g., $\theta \approx \pi/4$ and $3\pi/4$). Notably, it drops at the equator ($\theta = \pi/2$) because the deterministic force driving the oscillator scales with $\sin(\theta)$; its derivative, which governs the parameter sensitivity, vanishes for equatorial states.

Conversely, the sensitivity to the azimuthal angle, $\mathcal{I}_{\varphi\varphi}$ [Fig.~\ref{fig:fisher_maps}(b)], reaches its absolute maximum precisely at the equator, where an equal superposition state maximizes the amplitude of the quantum-induced non-stationary noise and deterministic force. Moreover, we emphasize a stark asymmetry in overall magnitude: $\mathcal{I}_{\theta\theta}$ is several orders of magnitude larger than $\mathcal{I}_{\varphi\varphi}$, reflecting the intrinsic difficulty of extracting phase information. As expected, $\mathcal{I}_{\varphi\varphi}$ vanishes completely at the poles ($\theta = 0, \pi$). Fundamentally, this limited sensitivity arises because the dependence on $\varphi$ enters solely as a phase shift within the cosine correlation of the noise, hindering the continuous accumulation of information.

Finally, the cross-information $\mathcal{I}_{\theta\varphi}$ [Fig.~\ref{fig:fisher_maps}(c)] vanishes along the equator, indicating that for maximal superposition states, $\theta$ and $\varphi$ can be estimated independently. At intermediate latitudes, non-zero cross-information introduces estimation correlations, which compromise simultaneous multiparameter reconstruction. Consequently, optimal parameter estimation is achieved precisely when the qubit resides in a superposition state, highlighting a fundamental metrological principle: the stronger the quantum coherence of the system, the greater the amount of extractable information.

\begin{figure}
    \centering
    \includegraphics[width=8.6cm]{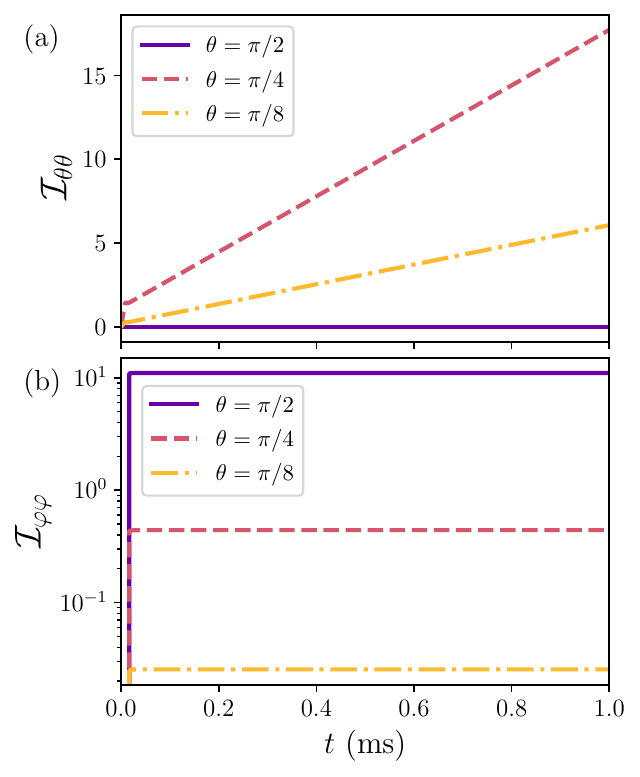}
    \caption{Temporal evolution of the Fisher information for continuous monitoring of the classical oscillator. (a) Information regarding the initial population, $\mathcal{I}_{\theta\theta}$, grows continuously with the integration time $t$. The instantaneous non-zero value at $t \to 0$ arises from the population's direct encoding into the amplitude of the initial deterministic force. (b) Information concerning the azimuthal phase, $\mathcal{I}_{\varphi\varphi}$, strictly vanishes at $t=0$ and saturates rapidly as the quantum superposition decays. The curves correspond to three representative initial states: maximal superposition ($\theta = \pi/2$, red solid line), intermediate asymmetry ($\theta = \pi/4$, orange dashed line), and high polar localization ($\theta = \pi/8$, blue dash-dotted line). In all cases, $\varphi = 0$. The parameters are chosen to be $r = 0.1$, $g = 0.1$, and $\Gamma = 10^{-3}$.}
    \label{fig:fisher over time}
\end{figure}

Next we examine the gain of information over time. Setting the initial phase to $\varphi = 0$ without loss of generality (as its dependency merely appears as a phase shift in the oscillatory functions), Fig. \ref{fig:fisher over time} reveals a fundamental asymmetry in the parameter estimation dynamics for the population $\theta$ and the phase $\varphi$. As shown in Fig. \ref{fig:fisher over time}(b), the information concerning the azimuthal phase, $\mathcal{I}_{\varphi\varphi}$, saturates rapidly. Because the phase $\varphi$ is exclusively encoded in the non-stationary quantum fluctuations and the deterministic force, it is exponentially suppressed by the qubit's decoherence. Once the quantum superposition decays, no further phase information can be extracted from the classical oscillator.

Conversely, Fig. \ref{fig:fisher over time}(a) demonstrates that the information regarding the population, $\mathcal{I}_{\theta\theta}$, grows linearly with the integration time. This striking behavior occurs because the initial population $\theta$ is also imprinted onto the stationary background noise of the oscillator. By continuously monitoring the classical trajectory, one accumulates statistics of this stationary noise variance, allowing the persistent extraction of population information long after the initial quantum coherence has vanished.

Moreover, the instantaneous jump of $\mathcal{I}_{\varphi\varphi}$ to a finite value immediately after $t=0$ reflects the fact that the azimuthal angle $\varphi$ dictates the initial phase shift of the transient quantum-induced interaction. Conversely, the information for the population angle, $\mathcal{I}_{\theta\theta}$, strictly vanishes at $t=0$ and grows continuously over time. This distinct temporal behavior arises because $\theta$ acts as a strength parameter governing the ongoing dynamics; therefore, extracting information about the population requires a finite integration window during which the oscillator maps out its trajectory, rendering the quantum effects classically distinguishable.

In summary, not all quantum states are equally estimable through classical noise spectroscopy. For the chosen parameters, the state that maximizes the information about the latitude is $(\theta \approx 0.979 \text{ rad}, \varphi = 2\pi)$, while the optimal state for phase estimation lies on the equator $(\theta = \pi/2, \varphi = n\pi)$. The points where $\varphi = n\pi$ represent preferential axes where the parameter correlation completely vanishes, providing an optimal configuration for quantum-to-classical transduction. Furthermore, the continuous temporal acquisition of information heavily favors $\theta$ over $\varphi$, a direct consequence of the azimuthal angle manifesting solely as a phase shift within the transient oscillatory dynamics.

\section{Conclusion}\label{sec: final remarks}




In this work, we have employed the Feynman-Vernon influence functional formalism to derive the effective classical dynamics of a mesoscopic oscillator coupled to a two-level quantum system via the Jaynes-Cummings interaction, since path integral allows us to bypass the use of operators for the oscillator. 
{This path integral approach provides a powerful organization structure, not only clarifying the physical interpretation, but also enabling to extract qualitative features of the effective dynamics without necessarily solving the equations of motion \cite{calzetta1994noise}. 
Moreover, the method employed is natural to understand the use of the classical system as a detector of quantum features, given that the influence of the external system depends on its inherently quantum properties, such as the qubit being in a superposition.

Our central finding is that the qubit induces both a state-dependent deterministic force and a stochastic noise source on the oscillator, with both stationary and non-stationary components. By quantifying the parameter estimation precision, we demonstrate that the continuous classical record acts as a powerful metrological tool. We reveal a temporal asymmetry: while information regarding the azimuthal phase $\varphi$ is strictly bounded by the decoherence timescale, the initial population $\theta$ becomes permanently imprinted onto the variance of the oscillator's stationary background noise. Consequently, information about $\theta$ can be persistently accumulated long after the initial quantum coherence has vanished. These quantum-induced forces, evaluated here for different spin-optomechanical systems \cite{meekhof1996generation,yin2013large, bild2023schrodinger}, imprint measurable signatures of the qubit's quantum state onto the oscillator's classical dynamics. Overcoming current limitations in phase estimation opens exciting pathways for future investigations. In this context, the exploration of different coupling architectures or time-modulated interactions could potentially map phase information onto the steady-state dynamics, further enhancing the metrological reach of classical probes in hybrid quantum systems.

The hybrid quantum-classical dynamics of a qubit and mechanical oscillator can find a number of interesting applications. For instance, by monitoring the classical oscillator, the qubit's state can be partially reconstructed, transducing quantum information into a classical signal. Moreover, the emergence of a deterministic quantum-induced force implies that the qubit exerts work on the mechanical oscillator, with potential implications to quantum stochastic thermodynamics and mesoscopic engines~\cite{peliti2021stochastic, paraguassu2025apparent}.

Our work also opens interesting avenues for future investigation. An immediate extension, for example, would be to analyze more complex systems such as an ensemble of qubits interacting with a mechanical oscillator~\cite{stamp1994dissipation, stamp2006decoherence}, which could be sensitive to collective entanglement-induced deterministic and stochastic effects, especially if the assumption of an initially separable spin-mechanical state is violated \cite{anglin1997deconstructing}. The investigation of different Hamiltonians, beyond the Jaynes-Cummings model, could also uncover new forms of quantum-induced forces \cite{martinetz2020quantum}.

Promising platforms for applications of our approach include hybrid systems where a levitated nanoparticle (the oscillator) is coupled to a trapped ion (the qubit) \cite{bykov2024nanoparticle}, as well as nanomechanical resonators coupled to solid-state two-level systems \cite{armour2002entanglement, oconnell2010quantum}. Finally, quantum-induced spin-dependent forces could find applications in spin-entanglement witnesses of the quantum nature  of gravity \cite{spin_collab}. 




\acknowledgments{
We acknowledge Nami F. Svaiter, Bruno Suassuna, Gabriel Audi and Amir Caldeira for useful conversations. P.V.P acknowledges the Funda\c{c}\~ao de Amparo \`a Pesquisa do Estado do Rio de Janeiro (FAPERJ Process SEI-260003/000174/2024). T.G. acknowledges the Coordena\c{c}\~ao de Aperfei\c{c}oamento de Pessoal de N\'ivel Superior - Brasil (CAPES) - Finance Code 001, Conselho Nacional de Desenvolvimento Cient\'ifico e Tecnol\'ogico (CNPq), Funda\c{c}\~ao de Amparo \`a Pesquisa do Estado do Rio de Janeiro (FAPERJ Scholarship No. E-26/200.252/2023 and E-26/202.762/2024)  and Funda\c{c}\~ao de Amparo \`a Pesquisa do Estado de São Paulo (FAPESP process No. 2021/06736-5). This work was supported by the Serrapilheira Institute (grant No. Serra – 2211-42299 ).}

\bibliography{bibliography.bib}

\newpage
\onecolumngrid
\appendix

\section{The BCH formula in the product of time ordered evolution operator}\label{appendixA}

With the dimensionless quantities, we expand our time ordered evolution operator, Eq.~\eqref{eq: atom time evolution operator}, using a Dyson series \cite{itzykson2006quantum}, until second order in the dimensionless coupling, i.e.,
\begin{equation}\label{eq: atom time evolution operator v2}
    U_{A,I}(X) \approx e^{-\frac{i}{\hbar}\int_{0}^{T}d\tau H_{\text{int},I}(\tau) - \frac{1}{2\hbar^2}\int_{0}^{T}d\tau\int_{0}^{\tau}d\tau' [H_{\text{int},I}(\tau),H_{\text{int},I}(\tau')]}, \nonumber
\end{equation}
discarding terms $\mathcal{O}(g^3)$. 

The interaction Hamiltonian with dimensionless parameters reads
\begin{equation}\label{eq: interaction Hamiltionian v2}
    \frac{H_{\text{int}}}{\hbar} = g \Big( \sigma_xf_x(\tau) - \sigma_yf_y(\tau) \Big),
\end{equation}
where $f_{x,y}(\tau)$ are given by Eqs \eqref{eq: fx} and \eqref{eq: fy}. Then the commutator $[H_{\text{int}}(\tau),H_{\text{int}}(\tau')]$ will be given by $$[H_{\text{int}}(\tau),H_{\text{int}}(\tau')] = 2ig^2\hbar^2 \sigma_z \big(f_x(\tau)f_y(\tau') - f_x(\tau')f_y(\tau) \big),$$ since $[\sigma_x,\sigma_y] = 2i\sigma_z$. Therefore, the evolution operators can be written as
\begin{eqnarray}
        U_{A,I}(X) &=& \exp \Big(-ig\sigma_xF_x+ig\sigma_yF_y - ig^2\sigma_zF_z + \mathcal{O}(g^3)\Big), \label{eq: atom time evolution operator vfinal} \\
        U_{A,I}^{\dagger}(X')
        &=& \exp \Big(+ig\sigma_xF'_x-ig\sigma_yF'_y + ig^2\sigma_zF'_z + \mathcal{O}(g^3)\Big), \label{eq: atom time evolution operator dagger vfinal}
    \end{eqnarray}
where
\begin{eqnarray}
    F_x &=& \int_{0}^{T}d\tau  f_x(\tau), \\
    F_y &=& \int_{0}^{T}d\tau f_y(\tau), \\
    F_z &=& \int_{0}^{T}d\tau\int_{0}^{\tau}d\tau' \Big( f_x(\tau)f_y(\tau') - f_x(\tau')f_y(\tau) \Big).
\end{eqnarray}

To finally be able to write down the influence functional, we invoke the Baker-Campbell-Hausdorff formula:
\begin{equation}\label{eq: BCH formula}
    e^{A}e^{B}\approx e^{A+B}e^{\frac{1}{2}[A,B]},
\end{equation}
which is valid because we are discarding terms $\mathcal{O}(g^3)$. For simplicity, let us define an operator $A$ such that $A = -i g \sigma_x F_x + i g \sigma_y F_y - i g^2 \sigma_z F_z$ while $B = A'^\dagger$. The commutator between both is given by:
\begin{equation}
    [A,B] = -2ig^2\sigma_z(F'_x F_y - F'_y F_x) + \mathcal{O}(g^3),
\end{equation}
while the sum of $A$ and $B$ gives us
\begin{equation}
    A+B = i g \sigma_x W_x + i g \sigma_y W_y + i g^2 \sigma_z (F'_z - F_z),
\end{equation}
where
\begin{eqnarray}
    W_x &=& F'_x - F_x, \\
    W_y &=& F_y - F'_y, \\
    W_z &=& F'_z + 2F'_xF_y - F'_xF'_y -F_xF_y - F_z.
\end{eqnarray}
Therefore, at the end of the day, we arrive at the following expression:
\begin{equation}
    e^{A}e^{B} = e^{i g \sigma_x W_x + i g \sigma_y W_y + i g^2 \sigma_z W_z}e^{ig^2\sigma_z(F'_x F_y - F'_y F_x) + \mathcal{O}(g^3)},
\end{equation}
and using the variant Baker-Campbell-Hausdorff formula again:
\begin{equation}
    e^{i g \sigma_x W_x + i g \sigma_y W_y + i g^2 \sigma_z W_z} = e^{i g \sigma_x W_x}e^{i g \sigma_y W_y}e^{i g^2 \sigma_z W_z}e^{-\frac{(ig)^2}{2}[\sigma_x W_x,\sigma_y W_y] + \mathcal{O}(g^3)}. 
\end{equation}
Finally we can combine these last equations and simplify which gives us the following:
\begin{equation}
    e^{A}e^{B} = e^{ig \sigma_x W_x}e^{ig \sigma_y W_y}e^{ig^2\sigma_z W_z}e^{\mathcal{O}(g^3)}.
\end{equation}
Note that it is exactly the product $U_{A,I}(T,0;X)U_{A,I}^{\dagger}(T,0;X')$, i.e.,
\begin{equation}\label{eq: atom time evolution operator product}
    U_{A,I}(X)U_{A,I}^{\dagger}(X')\approx e^{ig \sigma_x W_x}e^{ig \sigma_y W_y}e^{ig^2\sigma_zW_z}.
\end{equation}
Therefore, by insertion into Eq.~\eqref{eq: pre influence} we found the influence functional given in Eq.~\eqref{eq: influence functional version 1}.

\section{Calculating the influence functional}\label{appendixB}

Let us start by defining the already known Pauli operators:
\begin{equation}
    \sigma_x = \left( \begin{array}{cc}
        0 & 1 \\
        1 & 0
    \end{array} \right), ~\sigma_y = \left( \begin{array}{cc}
        0 & -i \\
        i & 0
    \end{array}\right), ~\sigma_z = \left( \begin{array}{cc}
        1 & 0 \\
        0 & -1
    \end{array} \right),
\end{equation}
and also the two qubit states:
\begin{equation}
    |0\rangle = \left( \begin{array}{c}
        1 \\
        0
    \end{array} \right), ~|1\rangle = \left( \begin{array}{c}
        0 \\
        1
    \end{array} \right).
\end{equation}
The influence functional $\mathcal{F}[X,X']$ is defined by Eq. \eqref{eq: influence functional} and the initial state of the qubit $| \Psi \rangle$ by Eq. \eqref{eq: atom inital state}. So, to evaluate this calculation and obtain a form of the influence functional, we need to understand how these operators coming from Eq. \eqref{eq: atom time evolution operator product} actuate in $|0\rangle$ and $|1\rangle$. The $\sigma_z$ is very simple and will only get a phase, i.e.,
\begin{equation}
    e^{ig^2 W_z \sigma_z}\ket{0} = e^{+ig^2 W_z}\ket{0}, ~e^{ig^2 W_z \sigma_z}\ket{1} = e^{-ig^2 W_z}\ket{1}, \label{eq: z operator}
\end{equation}
therefore, the influence functional reads
\begin{equation}\label{eq: influence functional v2}
    \mathcal{F}[X,X'] = \sqrt{1-p}~e^{+ig^2W_z}\bra{\Psi} e^{ig W_x\sigma_x }e^{ig  W_y\sigma_y}\ket{0} + e^{i\varphi}\sqrt{p}~e^{-ig^2W_z}\bra{\Psi} e^{ig W_x\sigma_x }e^{ig  W_y\sigma_y} \ket{1}.
\end{equation}
For these other two operators we can see explicitly. Using Taylor's series combined with the fact that $\sigma^{2}_k = \mathbf{1}$, we have:
\begin{equation}
    e^{ig W_{x,y}\sigma_{x,y}} = \mathbf{1}\cos(g W_{x,y}) + i \sigma_{x,y} \sin(g W_{x,y})
\end{equation}
so we get:
\begin{eqnarray}
    &&e^{ig W_{x}\sigma_{x}}\ket{0} = \cos(g W_x) \ket{0} + i \sin(g W_x)\ket{1}, ~e^{ig W_{x}\sigma_{x}}\ket{1} = \cos(g W_x) \ket{1} + i \sin(g W_x)\ket{0}, \label{eq: x operator} \\
    &&e^{ig W_{y}\sigma_{y}}\ket{0} = \cos(g W_y) \ket{0} - \sin(g W_y)\ket{1}, ~~~e^{ig W_{y}\sigma_{y}}\ket{1} = \cos(g W_y) \ket{1} + \sin(g W_y)\ket{0}. \label{eq: y operator}
\end{eqnarray}
Let us proceed step by step. First Eq. \eqref{eq: influence functional v2} becomes:
\begin{eqnarray}\label{eq: influence functional v3}
    \mathcal{F}[X,X'] 
    &=& \sqrt{1-p}~e^{+ig^2W_z}\Big(\cos(g W_y) \bra{\Psi} e^{ig W_x\sigma_x }\ket{0} - \sin(g W_y)\bra{\Psi} e^{ig W_x\sigma_x }\ket{1}\Big) \\
    && +e^{i\varphi}\sqrt{p}~e^{-ig^2W_z}\Big(\cos(g W_y) \bra{\Psi} e^{ig W_x\sigma_x }\ket{1} + \sin(g W_y)\bra{\Psi} e^{ig W_x\sigma_x }\ket{0}\Big), \nonumber
\end{eqnarray}
then we need to apply the remainder operator, this leads us to the following expression:
\begin{eqnarray}\label{eq: influence functional v4}
    \mathcal{F}[X,X'] 
    &=& \sqrt{1-p}~e^{+ig^2W_z}\Bigg(\cos(g W_y) \Big(\cos(g W_x)\langle\Psi|0\rangle + i \sin(g W_x)\langle \Psi | 1 \rangle\Big) \\
    && ~~~~~~~~~~~~~~~~~~~~~~  - \sin(g W_y)\Big(\cos(g W_x) \langle \Psi |1 \rangle + i \sin(g W_x)\langle \Psi | 0 \rangle\Big)\Bigg) \nonumber\\
    && +e^{i\varphi}\sqrt{p}~e^{-ig^2W_z}\Bigg(\cos(g W_y)\Big(\cos(g W_x) \langle \Psi | 1 \rangle + i \sin(g W_x)\langle \Psi | 0 \rangle\Big)\nonumber\\
    && ~~~~~~~~~~~~~~~~~~~~~~ + \sin(g W_y)\Big(\cos(g W_x)\langle \Psi | 0 \rangle + i \sin(g W_x)\langle \Psi | 1 \rangle\Big)\Bigg), \nonumber
\end{eqnarray}
rearranging
\begin{eqnarray}\label{eq: influence functional v5}
    \mathcal{F}[X,X'] 
    &=& \sqrt{1-p}~e^{+ig^2W_z} \Bigg(\langle\Psi|0\rangle\Big(\cos(g W_y)\cos(g W_x) - i \sin(g W_y)\sin(g W_x)\Big)  \\
    && ~~~~~~~~~~~~~~~~~~~~~~  - \langle \Psi |1 \rangle \Big(\sin(g W_y)\cos(g W_x)-i \cos(g W_y)\sin(g W_x) \Big)\Bigg) \nonumber\\
    && +e^{i\varphi}\sqrt{p}~e^{-ig^2W_z}\Bigg(\langle \Psi | 1 \rangle\Big(\cos(g W_y)\cos(g W_x) + i \sin(g W_y)\sin(g W_x) \Big)\nonumber\\
    && ~~~~~~~~~~~~~~~~~~~~~~ + \langle \Psi | 0 \rangle\Big( \sin(g W_y)\cos(g W_x)  + i \cos(g W_y)\sin(g W_x)\Big)\Bigg), \nonumber
\end{eqnarray}
and now replacing $\langle \Psi | 0 \rangle = \sqrt{1-p}$ and $\langle \Psi | 1 \rangle = e^{-i\varphi}\sqrt{p}$, we finally have
\begin{eqnarray}\label{eq: influence functional v6}
    \mathcal{F}[X,X'] 
    &=& e^{+ig^2W_z} \Bigg((1-p)\Big(\cos(g W_y)\cos(g W_x) - i \sin(g W_y)\sin(g W_x)\Big)  \\
    && ~~~~~~~~~~~~~~  - e^{-i\varphi}\sqrt{p(1-p)}\Big(\sin(g W_y)\cos(g W_x)-i \cos(g W_y)\sin(g W_x) \Big)\Bigg) \nonumber\\
    && +e^{-ig^2W_z}\Bigg(p\Big(\cos(g W_y)\cos(g W_x) + i \sin(g W_y)\sin(g W_x) \Big)\nonumber\\
    && ~~~~~~~~~~~~~~ + e^{i\varphi}\sqrt{p(1-p)}\Big( \sin(g W_y)\cos(g W_x)  + i \cos(g W_y)\sin(g W_x)\Big)\Bigg). \nonumber
\end{eqnarray}
This final expression is too complex for us to understand how it will generate an effective action in Eq. \eqref{eq: density matrix}. Therefore, we can express $\mathcal{F} = \exp(\log \mathcal{F})$ and, as $g \ll 1$, expand $\log \mathcal{F}$ in powers of $g$. We find
\begin{eqnarray}
    \log \mathcal{F}[X,X'] 
    &=& 2ig\sqrt{p(1-p)}\Big(W_x\cos\varphi + W_y \sin\varphi \Big) + i g^2(1-2p)\big(W_z - W_x W_y\big) \\
    && - \frac{g^2}{2}\Big(W^2_x + W^2_y - 2p(1-p)\big( W^2_x(1+\cos(2\varphi)) + W^2_y(1-\cos(2\varphi)) + 2 W_x W_y \sin(2\varphi) \big)  \Big) + \mathcal{O}(g^3). \nonumber
\end{eqnarray}

These calculations were also performed using Mathematica \cite{mathematica} and can be seen in the Supplemental Material.

\section{Heisenberg equations}

Let us now analyze the use of a different method, namely the Heisenberg approach. The Heisenberg method is very commonly used in quantum optics, and is based on writing equations of motion for creation and annihilation operators $a, a^\dagger $ and qubit observables $ \sigma_i$. As it turns out, these operators obey a system of coupled ordinary differential equations (ODEs). For the case of two linearly coupled oscillators, it turns out the Heisenberg approach is the easiest way to derive this kind of quantum-induced stochastic dynamics studied in the current work. For the case of interest here, however, it becomes significantly more involved, as detailed in the following. The path integral method, on the other is directly amenable to perturbation theory, making its application straightforward.

For the case of interest in the current manuscript, the coupled system of ODEs read,
\begin{align}
    \dot{q} &= \omega_o p - \frac{\Omega}{2\sqrt{2}}\sigma_y, ~
    \dot{p} = -\omega_o q - \frac{\Omega}{2\sqrt{2}}\sigma_x, \label{eq: oscillator} \\
    \dot{\sigma}_x &= - \omega_q \sigma_y -\frac{\Omega}{\sqrt{2}} p \sigma_z, ~
    \dot{\sigma}_y = \omega_q \sigma_x-\frac{\Omega}{\sqrt{2}} q \sigma_z, ~
    \dot{\sigma}_z = \frac{\Omega}{\sqrt{2}} \big( q\sigma_y + p\sigma_x \big) \label{eq: qubit}
\end{align}

{We may then write Eq.\eqref{eq: qubit} in matrix form,
\begin{eqnarray}
    \frac{d}{dt}\boldsymbol{\sigma}(t) = \tilde{H}(t)\boldsymbol{\sigma}(t), \label{eq: heisenberg}
\end{eqnarray}
where $\tilde{H}(t)$ is a time-dependent Hamiltonian, which also depends on the oscillator observables $q(t)$ and $p(t)$, and $\boldsymbol{\sigma}(t)$ = $(\sigma_x,\sigma_y,\sigma_z)^{T}$. In order to solve this system of equations, we have to introduce the time ordering operator in a Dyson series, which will lead to very similar steps as in the path integral formalism; very similar steps arise in the path integral formulation --  we have some sort of conservation of difficulty.}

Using Eq.~\eqref{eq: heisenberg}, we begin by introducing the dimensionless time
\(\tau = \omega_q t\). We also define the dimensionless parameters
\(2\sqrt{2}\,g = \Omega/\omega_q\) and \(r = \omega_o/\omega_q\).
With these definitions, the time-dependent Hamiltonian can be written as
\begin{equation}
    \tilde{H}(\tau) = \tilde{H}^{(0)} + 2g\,\tilde{H}^{(1)}(\tau),
\end{equation}
represented by the matrices
\begin{equation}
    [\tilde{H}^{(0)}] =
    \begin{pmatrix}
        0 & -1 & 0 \\
        1 & 0  & 0 \\
        0 & 0  & 0
    \end{pmatrix},
    \qquad
    [\tilde{H}^{(1)}(\tau)] =
    \begin{pmatrix}
        0 & 0 & -p(\tau) \\
        0 & 0 & -q(\tau) \\
        p(\tau) & q(\tau) & 0
    \end{pmatrix}.
\end{equation}

The solution to Eq.~\eqref{eq: heisenberg} can be expressed as
\begin{equation}
    \boldsymbol{\sigma}(\tau) = U(\tau)\,\boldsymbol{\sigma}(0),
\end{equation}
where the evolution operator \(U(\tau)\) satisfies
\begin{equation}
    \frac{d}{d\tau}U(\tau) = \tilde{H}(\tau)\,U(\tau),
    \qquad
    U(0) = I.
\end{equation}

We assume a perturbative expansion in the coupling constant \(g\),
\begin{equation}
    U(\tau) = U^{(0)}(\tau) + g\,U^{(1)}(\tau) + \mathcal{O}(g^2),
\end{equation}
which leads to the following hierarchy of equations:
\begin{eqnarray}
    \frac{d}{d\tau}U^{(0)}(\tau)
    &=&
    \tilde{H}^{(0)}\,U^{(0)}(\tau),
    \\
    \frac{d}{d\tau}U^{(1)}(\tau)
    &=&
    \tilde{H}^{(0)}\,U^{(1)}(\tau)
    + 2\,\tilde{H}^{(1)}(\tau)\,U^{(0)}(\tau).
\end{eqnarray}
The corresponding solutions are given by
\begin{eqnarray}
    U^{(0)}(\tau)
    &=&
    \exp\!\left(\tilde{H}^{(0)}\,\tau\right),
    \\
    U^{(1)}(\tau)
    &=&
    2\,U^{(0)}(\tau)
    \int_0^\tau d\tau'\,
    \bigl[U^{(0)}(\tau')\bigr]^{-1}
    \tilde{H}^{(1)}(\tau')\,
    U^{(0)}(\tau').
\end{eqnarray}

One can straightforwardly verify that
\begin{equation}
    U^{(0)}(\tau)
    \equiv R(\tau)
    =
    \begin{pmatrix}
        \cos\tau & -\sin\tau & 0 \\
        \sin\tau & \cos\tau  & 0 \\
        0 & 0 & 1
    \end{pmatrix},
\end{equation}
which implies \([U^{(0)}(\tau)]^{-1} = R(-\tau)\).
A direct calculation of
$U^{(1)}$ then yields
\begin{eqnarray}
    \sigma_x(\tau)
    &=&
    \sigma_x(0)\cos\tau
    - \sigma_y(0)\sin\tau
    - 2g\,\sigma_z(0)
    \int_0^\tau d\tau'\,
    \bigl[
        q(\tau')\sin(\tau' - \tau)
        + p(\tau')\cos(\tau' - \tau)
    \bigr],
    \\
    \sigma_y(\tau)
    &=&
    \sigma_x(0)\sin\tau
    + \sigma_y(0)\cos\tau
    - 2g\,\sigma_z(0)
    \int_0^\tau d\tau'\,
    \bigl[
        q(\tau')\cos(\tau' - \tau)
        - p(\tau')\sin(\tau' - \tau)
    \bigr].
\end{eqnarray}

Substituting these expressions into Eq. \eqref{eq: oscillator},
and denoting derivatives with respect to \(\tau\) by dots, we obtain
\begin{eqnarray}
    \dot{q}(\tau)
    &=&
    r\,p(\tau)
    - g\bigl[
        \sigma_x(0)\sin\tau
        + \sigma_y(0)\cos\tau
    \bigr]
    + \mathcal{O}(g^2),
    \\
    \dot{p}(\tau)
    &=&
    -r\,q(\tau)
    - g\bigl[
        \sigma_x(0)\cos\tau
        - \sigma_y(0)\sin\tau
    \bigr]
    + \mathcal{O}(g^2).
\end{eqnarray}

Since these quantities are operators, we take expectation values with respect
to the initial state
\(\ket{\Psi} \otimes \ket{\psi_{\mathrm{osc}}}\),
where the qubit state \(\ket{\Psi}\) is given by Eq.~(24),
\begin{equation}\label{eq: atom initial state}
    \ket{\Psi}
    =
    \sqrt{1-p}\,\ket{0}
    + e^{i\varphi}\sqrt{p}\,\ket{1}.
\end{equation}
This yields
\begin{equation}
    \bra{\Psi}\sigma_x(0)\ket{\Psi}
    = 2\sqrt{p(1-p)}\cos\varphi,
    \qquad
    \bra{\Psi}\sigma_y(0)\ket{\Psi}
    = 2\sqrt{p(1-p)}\sin\varphi,
\end{equation}
while we define the classical oscillator variables as
\(q_{\mathrm{cl}}(\tau)
= \bra{\psi_{\mathrm{osc}}}q(\tau)\ket{\psi_{\mathrm{osc}}}\)
and
\(p_{\mathrm{cl}}(\tau)
= \bra{\psi_{\mathrm{osc}}}p(\tau)\ket{\psi_{\mathrm{osc}}}\).
Consequently, we arrive at
\begin{eqnarray}
    \dot{q}_{\mathrm{cl}}(\tau)
    &=&
    r\,p_{\mathrm{cl}}(\tau)
    - 2g\sqrt{p(1-p)}\sin(\tau + \varphi)
    + \mathcal{O}(g^2),
    \\
    \dot{p}_{\mathrm{cl}}(\tau)
    &=&
    -r\,q_{\mathrm{cl}}(\tau)
    - 2g\sqrt{p(1-p)}\cos(\tau + \varphi)
    + \mathcal{O}(g^2),
\end{eqnarray}
which correspond exactly to Eqs.\eqref{eq:q_final} and \eqref{eq:p_final} in the absence of noise terms.

\section{Regularizing the kernel by summing over modes}\label{appendixD}

In the main text we showed that the equations of motion for the quadrature of the qubit reads
\begin{equation}
    m\frac{d^2x}{dt^2} + m\omega^2_o x  = -f(t;\theta,\varphi) + \eta(t),
\end{equation}
{where the force is given by}
\begin{equation}
    f(t;\theta,\varphi) = 2\mathbf{f}_0 \left( \frac{1-r}{r} \right)\eta_f\cos(\tau+\varphi) \equiv 2\mathbf{f}_0 \left( \frac{1-r}{r} \right) f_0(t;\theta,\varphi)
\end{equation}
and the noise $\eta(\tau)$ is satisfying 
\begin{equation}
    \langle \eta(t) \eta(t') \rangle = \frac{\mathbf{f}^2_0}{r^2}\Big\{ \left[2r+(1-r+r^2)\eta_f^2 \right]\cos(\tau-\tau') +\left[2(1-r^2)\eta^2_{st}-r \eta^2_f \right]\cos(\tau+\tau'+2\varphi)\Big\} \equiv \frac{\mathbf{f}^2_0}{r^2} \langle \eta(t) \eta(t') \rangle_0 
\end{equation}
{where $\eta_f(\theta)$ and $\eta_{st}(\theta)$ are given by Eqs.\eqref{eq: force intensity_theta} and \eqref{eq: force intensity2_theta} respectively, and }$r = \frac{\omega_o}{\omega_q}$. 
In the main text derivation we assumed a ``perfect" qubit, in which the transition frequency is precisely $\omega_q$. However, in reality, there is coupling of the qubit with the environment and even some imperfections in the physical qubit\cite{cywinski2008enhance, yuge2011measurement}, leading to a small noise in the transition frequency. To phenomenologically deal with that, we will broaden the transition frequency to a Lorentzian distribution centered around $\omega_q$,    
\begin{equation}
    J(\omega) =   \frac{1}{\pi} \frac{\Gamma / 2 }{(\omega - \omega_q)^2 + \left(\frac{\Gamma}{2}\right)^2}
\end{equation}
where the regime of interest is when $\omega_q \gg \Gamma$, meaning that the Lorentzian is sharply peaked around $\omega_q$.  

To invert the kernel, we then need to sum over the modes in the deterministic force generated by the qubit
\begin{equation}
    2\mathbf{f}_0 \left( \frac{1-r}{r} \right)\int_0^\infty d\omega ~ J(\omega) f_0(t;\theta,\varphi)
\end{equation}
and in the noise correlation function
\begin{equation}\label{eq:real_deal}
    \frac{\mathbf{f}^2_0}{r^2} \int_0^\infty d\omega ~J(\omega)  \langle \eta(t) \eta(t') \rangle_0
\end{equation}

Before diving in the calculations, note that
\begin{equation}
    \int_{-\infty}^{\infty} d\omega ~J(\omega) = 1,
\end{equation}
and
\begin{equation}
    \int_{0}^{\infty} d\omega ~J(\omega) = \frac{1}{2} + \frac{1}{\pi}\tan^{-1}\left(\frac{2\omega_q}{\Gamma}\right) = 1 - \mathcal{O}(\Gamma/\omega_q).
\end{equation}
Since we are considering only small variations from a perfect qubit, we have $\omega_q \gg \Gamma$ and the spectrum $J(\omega)$ is sharply peaked around $\omega_q$ with a small width. This implies that the contribution from frequencies $\omega \in (-\infty,0]$ is negligible in this regime. Moreover, considering the full frequency range is mathematically more consistent, as it allows for a better definition of contour integrals and avoids issues arising in terms such as $\omega^{-1}J(\omega)$ and $\omega^{-2}J(\omega)$, which exhibit logarithmic divergences as $\omega \to 0$.

For the reasons discussed earlier, let us rewrite Eq. \eqref{eq:real_deal}{using $r = \omega_o/\omega_q$ and $\tau = \omega_q t$} in order to highlight what kind of integration method we will use and also changing the integration limits 
\begin{equation}
    \int_{-\infty}^{\infty}d\omega J(\omega) \left\{ \underbrace{A_\theta \cos\omega \xi - B_\theta \cos2(\omega \eta+\varphi)}_{\text{term 1}} + 2\omega_o \underbrace{\omega^{-1} \left[ \cos\omega \xi - B_\theta \cos2(\omega \eta+\varphi) \right]}_{\text{term 2}} + \omega_o^2 \underbrace{\omega^{-2} \left[A_\theta \cos\omega \xi - B_\theta \cos2(\omega \eta+\varphi)\right]}_{\text{term 3}} \right\}, \label{eq:all_terms}
\end{equation}
where we defined a few quantities:
\begin{equation}
    A_\theta = \frac{1 + \cos^2\theta}{2}, B_\theta = \sin^2(\theta/2), ~\xi = t - t', ~\eta = (t+t')/2.
\end{equation}
Finally, let us solve each term by turn.
\subsection{Term 1}
 The term proportional to $\omega^0$ can be done without contour integration. First, we have:
\begin{align}
    \int_{-\infty}^{\infty}d\omega ~J(\omega)\cos \omega x &= \cos\omega_q x \left( \cosh\frac{\Gamma x}{2} - \frac{x}{|x|}\sinh\frac{\Gamma x}{2} \right), \text{ for x $\in \mathbb{R}$}, \\
    \int_{-\infty}^{\infty}d\omega ~J(\omega)\sin \omega x &= \sin\omega_q x \left( \cosh\frac{\Gamma x}{2} - \frac{x}{|x|}\sinh\frac{\Gamma x}{2} \right), \text{ for x $\in \mathbb{R}$}.
\end{align}
Since $\cos(a+b) = \cos a \cos b - \sin a \sin b$, we have:
\begin{align}
    \int_{-\infty}^{\infty} d\omega~J(\omega) \left[ A_\theta \cos\omega \xi - B_\theta \cos2(\omega \eta+\varphi) \right] &= A_\theta \cos\omega_q \xi \left( \cosh\frac{\Gamma \xi}{2} - \frac{\xi}{|\xi|}\sinh\frac{\Gamma \xi}{2} \right) \nonumber \\
    &- B_\theta \cos2(\omega_q \eta + \varphi) \left( \cosh\Gamma \eta - \frac{\eta}{|\eta|}\sinh\Gamma \eta \right). \nonumber
\end{align}
Using the relations 
\begin{equation*}
    \left( \cosh\frac{\Gamma \xi}{2} - \frac{\xi}{|\xi|}\sinh\frac{\Gamma \xi}{2} \right) = e^{-\frac{\Gamma |\xi|}{2}},\,\,\, \left( \cosh\Gamma \eta - \frac{\eta}{|\eta|}\sinh\Gamma \eta \right)= e^{-\Gamma|\eta|}, 
\end{equation*}
we have
\begin{align}
    \int_{-\infty}^{\infty} d\omega~J(\omega) \left[ A_\theta \cos\omega \xi - B_\theta \cos2(\omega \eta+\varphi) \right] &= A_\theta e^{-\Gamma |\xi|/2} \cos\omega_q \xi - B_\theta e^{-\Gamma \eta} \cos2(\omega_q \eta + \varphi),\label{eq:term1_final}
\end{align}
where $\eta > 0$.

\subsection{Term 2}

 We begin the second term by expanding $\cos x = \frac{1}{2}\left( e^{ix} + e^{-ix}\right)$ and solve each 4 terms by turn, we have:
\begin{equation}
    2\omega_o \int_{-\infty}^{\infty} d\omega ~J(\omega) \omega^{-1} \left[ \cos\omega \xi - B_\theta \cos2(\omega \eta+\varphi) \right] = \omega_0 \int_{-\infty}^{\infty} d\omega ~\omega^{-1}J(\omega)\left[ e^{i\omega \xi} + e^{-i\omega \xi} - B_\theta \left(e^{2i\varphi} e^{2i\omega \eta} + e^{-2i\varphi} e^{-2i\omega \eta} \right) \right]. \label{eq:term2_detailed}
\end{equation}
We have a trivial pole at $\omega = 0$ and $J(\omega)$ gives us two more: $\omega_p = \omega_q + i \Gamma/2$ and $\omega_p^* = \omega_q - i \Gamma/2$. Now, we can calculate the principal value of each term, let us define a function $f_\pm(\omega)$ given by
\begin{equation}
    f_\pm(\omega) = \frac{1}{\pi}\frac{\Gamma/2}{(\omega - \omega_p)(\omega - \omega_p^*)}\frac{e^{\pm i \omega z}}{\omega},
\end{equation}
computing the residue at each pole, we have
\begin{equation}\label{eq:residues}
    \text{Res}(f_\pm,0) = J(0) \equiv \frac{1}{\pi} \frac{\Gamma/2}{\omega_p \omega_p^*}, ~
    \text{Res}(f_\pm,\omega_p) = \frac{1}{\pi}\frac{\Gamma/2}{\omega_p -\omega_p^*}\frac{e^{\pm i\omega_p z}}{\omega_p}, ~
    \text{Res}(f_\pm,\omega_p^*) = \frac{1}{\pi}\frac{\Gamma/2}{\omega_p^* -\omega_p}\frac{e^{\pm i\omega_p^* z}}{\omega_p^*},
\end{equation}
where $\omega_p - \omega^*_p = i \Gamma$ and $|\omega_p|^2 = \omega_q^2 + (\Gamma/2)^2$. Let us proceed for each calculation.

We will start with the terms accompanying $e^{i \omega \xi}$ and $e^{2i \omega \eta}$.

Considering the integral $P.V. \int_{-\infty}^{\infty} f_+(\omega) d\omega$, the exponential behavior in the complex plane $(\omega = \omega_R + i \omega_I)$ is $|e^{i \omega z}| = e^{-\omega_I z}$. Therefore: 
for $z > 0$ we need $\omega_I > 0$ (upper half contour), for $z < 0$ we need $\omega_I < 0$ (lower half contour). Finally, we have:
\begin{align}
    P.V. \int_{-\infty}^{\infty} d\omega ~f_+(\omega) &= i \pi \bigg[ \Big( 2 \text{Res}(f_+,\omega_p) + J(0) \Big) \Theta(z) -\Big(2 \text{Res}(f_+,\omega_p^*) + J(0) \Big)\Theta(-z) \bigg]_{z=\xi}, \label{eq:term2_a}\\
    P.V. \int_{-\infty}^{\infty} d\omega ~f_+(\omega) &= i \pi \Big( 2 \text{Res}(f_+,\omega_p) + J(0) \Big) \Theta(z)\Bigg|_{z=2\eta}, \label{eq:term2_c}
\end{align}
and since $\eta = 2(t + t')$  can only be positive. Using Eq.\eqref{eq:residues} is easy to note that:
\begin{align}
    2\text{Res}(f_{\pm},\omega_p) + J(0) &= \frac{1}{\pi} \frac{\Gamma/2}{\omega_p} \left( \frac{2e^{\pm i \omega_p z}}{\omega_p - \omega_p^*} + \frac{1}{\omega_p^*} \right), \label{eq:residue_omega_p}\\
    2\text{Res}(f_{\pm},\omega_p^*) + J(0) &= \frac{1}{\pi} \frac{\Gamma/2}{\omega_p^*} \left( \frac{2e^{\pm i \omega_p^* z}}{\omega_p^* - \omega_p} + \frac{1}{\omega_p} \right). \label{eq:residue_omega_p*}
\end{align}
Therefore, Eqs.\eqref{eq:term2_a} and \eqref{eq:term2_c} become
\begin{align}
    P.V. \int_{-\infty}^{\infty} d\omega ~\omega^{-1} J(\omega) e^{i \omega \xi} &= i \bigg[ \frac{\Gamma/2 }{\omega_p} \left( \frac{2e^{i \omega_p \xi}}{\omega_p - \omega_p^*} + \frac{1}{\omega_p^*} \right) \Theta(\xi) -\frac{\Gamma/2 }{\omega_p^*} \left( \frac{2e^{i \omega_p^* \xi}}{\omega_p^* - \omega_p} + \frac{1}{\omega_p} \right)\Theta(-\xi) \bigg], \\
    P.V. \int_{-\infty}^{\infty} d\omega ~\omega^{-1} J(\omega) e^{2i \omega \eta} &= i  \frac{\Gamma/2}{\omega_p} \left( \frac{2e^{2i \omega_p \eta}}{\omega_p - \omega_p^*} + \frac{1}{\omega_p^*} \right) \Theta(\eta).
\end{align}

Next we proceed to compute the terms with negative exponential $e^{-i\omega \xi}$ and $e^{-2i\omega \eta}$.

Taking into account the integral $P.V. \int_{-\infty}^{\infty} f_-(\omega) d\omega$, the exponential behavior in the complex plane $(\omega = \omega_R + i \omega_I)$ is $|e^{-i\omega z}| = e^{i \omega_I z}$. Therefore: for $z > 0$ we need $\omega_I < 0$ (lower half contour), for $z < 0$ we need $\omega_I > 0$ (upper half contour). That gives us
\begin{align}
    P.V. \int_{-\infty}^{\infty} d\omega ~f_-(\omega) &= i \pi \bigg[ \Big( 2 \text{Res}(f_-,\omega_p) + J(0) \Big) \Theta(-z) -\Big(2 \text{Res}(f_-,\omega_p^*) + J(0) \Big)\Theta(z) \bigg]_{z=\xi}, \label{eq:term2_b}\\
    P.V. \int_{-\infty}^{\infty} d\omega ~f_-(\omega) &= -i \pi \Big( 2 \text{Res}(f_-,\omega_p^*) + J(0) \Big) \Theta(z)\Bigg|_{z=2\eta}, \label{eq:term2_d}
\end{align}
since $\eta > 0$ as discussed previously. Using Eqs.\eqref{eq:residue_omega_p} and \eqref{eq:residue_omega_p*} in Eqs.\eqref{eq:term2_b} and \eqref{eq:term2_d}, we obtain
\begin{align}
    P.V. \int_{-\infty}^{\infty} d\omega ~f_-(\omega) &= i \bigg[ \frac{\Gamma/2 }{\omega_p} \left( \frac{2e^{-i \omega_p \xi}}{\omega_p - \omega_p^*} + \frac{1}{\omega_p^*} \right) \Theta(-\xi) -\frac{\Gamma/2 }{\omega_p^*} \left( \frac{2e^{-i \omega_p^* \xi}}{\omega_p^* - \omega_p} + \frac{1}{\omega_p} \right)\Theta(\xi) \bigg]\\
    P.V. \int_{-\infty}^{\infty} d\omega ~f_-(\omega) &= -i \frac{\Gamma/2 }{\omega_p^*} \left( \frac{2e^{-2i \omega_p^* \eta}}{\omega_p^* - \omega_p} + \frac{1}{\omega_p} \right) \Theta(\eta), \label{eq:term2_d}
\end{align}

We can finally combine these 4 terms according to Eq. \eqref{eq:term2_detailed}. Let us do it by parts:
\begin{align}
    P.V.\int_{-\infty}^{\infty} d\omega ~\omega^{-1}J(\omega) \left( e^{i\omega \xi} + e^{-i \omega \xi} \right) &= i \Gamma/2 \bigg\{ \left[\frac{1}{\omega_p} \left( \frac{2e^{i \omega_p \xi}}{\omega_p - \omega_p^*} + \frac{1}{\omega_p^*} \right)-\frac{1}{\omega_p^*} \left( \frac{2e^{-i \omega_p^* \xi}}{\omega_p^* - \omega_p} + \frac{1}{\omega_p} \right) \right]\Theta(\xi) \nonumber \\
    & ~~~~~~~~~+ \left[ \frac{1}{\omega_p} \left( \frac{2e^{-i \omega_p \xi}}{\omega_p - \omega_p^*} + \frac{1}{\omega_p^*} \right) -\frac{1}{\omega_p^*} \left( \frac{2e^{i \omega_p^* \xi}}{\omega_p^* - \omega_p} + \frac{1}{\omega_p} \right) \right] \Theta(-\xi) \bigg\} \nonumber \\
    &= \frac{i \Gamma}{\omega_p - \omega_p^*} \bigg\{ \left(\frac{e^{i \omega_p \xi}}{\omega_p} +\frac{e^{-i \omega_p^* \xi}}{\omega_p^*}\right)\Theta(\xi) + \left( \frac{e^{-i \omega_p \xi}}{\omega_p}  +\frac{e^{i \omega_p^* \xi}}{\omega_p^*}\right) \Theta(-\xi) \bigg\}, \nonumber
\end{align}
but one can remember that $\omega_p = \omega_q + i \Gamma/2$ and use the well-known relation $z + z^* = 2 \text{Re}(z)$, to obtain
\begin{align}
    P.V.\int_{-\infty}^{\infty} d\omega ~\omega^{-1}J(\omega) \left( e^{i\omega \xi} + e^{-i \omega \xi} \right) &= 2\text{Re}\left[ \frac{e^{i\omega_p \xi} \Theta(\xi) + e^{-i\omega_p \xi} \Theta(-\xi)}{\omega_p} \right] \nonumber \\
    &= 2e^{- \Gamma|\xi|/2 } \left(\frac{\omega_q \cos\omega_q |\xi| + (\Gamma/2) \sin\omega_q |\xi|}{\omega_q^2 + (\Gamma/2)^2}\right).
\end{align}

The other term is
\begin{align}
    P.V.\int_{-\infty}^{\infty} d\omega ~\omega^{-1}J(\omega) \left( e^{2i\varphi}e^{2i\omega \eta} + e^{-2i\varphi}e^{-2i \omega \eta} \right) &= i \Gamma/2 \bigg[ \frac{e^{2i\varphi}}{\omega_p} \left( \frac{2e^{2i \omega_p \eta}}{\omega_p - \omega_p^*} + \frac{1}{\omega_p^*} \right) - \frac{e^{-2i\varphi}}{\omega_p^*} \left( \frac{2e^{-2i \omega_p^* \eta}}{\omega_p^* - \omega_p} + \frac{1}{\omega_p} \right)  \bigg]\Theta(\eta) \nonumber\\
    & = i \Gamma/2 \bigg[ \frac{2}{\omega_p - \omega_p^*} \left( \frac{e^{2i(\omega_p \eta + \varphi)}}{\omega_p} + \frac{e^{-2i(\omega_p^* \eta + \varphi)}}{\omega_p^*}\right) + \frac{e^{2i\varphi} - e^{-2i\varphi}}{\omega_p \omega_p^*}  \bigg]\Theta(\eta), \nonumber
\end{align}
using the well-known relation $e^{iz} - e^{-iz} = 2i \sin z$, we can obtain:
\begin{align}
    P.V.\int_{-\infty}^{\infty} d\omega ~\omega^{-1}J(\omega) \left( e^{2i\varphi}e^{2i\omega \eta} + e^{-2i\varphi}e^{-2i \omega \eta} \right) &= \left\{ 2 \text{Re}\left[ \frac{e^{2i(\omega_p \eta + \varphi)}}{\omega_p} \right] - \frac{\Gamma}{|\omega_p|^2} \sin2\varphi \right\}\Theta(\eta) \nonumber \\
    & = 2 e^{-\Gamma \eta} \left( \frac{\omega_q \cos2(\omega_q \eta + \varphi) + (\Gamma/2) \sin2(\omega_q \eta + \varphi)}{\omega_q^2 + (\Gamma/2)^2} \right) - \frac{\Gamma \sin2\varphi}{\omega^2_q + (\Gamma/2)^2 },
\end{align}
where $\eta > 0$.

Finally, the complete expression is given by:
\begin{align}
    &2\omega_o \int_{-\infty}^{\infty} d\omega ~J(\omega) \omega^{-1} \left[ \cos\omega \xi - B_\theta \cos2(\omega \eta+\varphi) \right] \label{eq:term2_final}\\
    &= \frac{2\omega_o}{\omega_q^2 + (\Gamma/2)^2} \left\{ e^{- \Gamma|\xi|/2 } \left(\omega_q \cos\omega_q \xi + (\Gamma/2) \sin\omega_q |\xi|\right) -  B_\theta \left[ e^{-\Gamma \eta} \left( \omega_q \cos2(\omega_q \eta + \varphi) + (\Gamma/2) \sin2(\omega_q \eta + \varphi) \right) - (\Gamma/2) \sin2\varphi \right] \right\}.\nonumber
\end{align}

In the limit $\omega_q\gg \Gamma$ we have
\begin{equation}
    2\omega_o \int_{-\infty}^{\infty} d\omega ~J(\omega) \omega^{-1} \left[ \cos\omega \xi - B_\theta \cos2(\omega \eta+\varphi) \right] \approx 2\left(\frac{\omega_o}{\omega_q}\right) \left\{ e^{- \Gamma|\xi|/2 }  \cos\omega_q \xi  -  B_\theta  e^{-\Gamma \eta}  \cos2(\omega_q \eta + \varphi) \right\}.
\end{equation}

\subsection{Term 3}
The procedure for the third term is analogous as before. The only difference comes from the fact that now we have four poles, instead of three, which comes from the fact that the $\omega =0$ is now a second order pole. 
By defining
\begin{equation}
    F_\pm(\omega) = \frac{1}{\pi} \frac{\Gamma/2}{(\omega-\omega_p)(\omega - \omega_p^*)}\frac{e^{\pm i \omega z}}{\omega^2},
\end{equation}
we get to a very similar calculation as done inn the previous term, simply replacing $f_\pm \to F_\pm$. Collecting all the terms, eventually we end up with
\begin{align}
    &\omega_o^2 \int_{-\infty}^{\infty}d\omega~J(\omega)\omega^{-2}\left[ A_\theta \cos\omega\xi - B_\theta \cos2(\eta+\varphi) \right] \label{eq:term3_final}\\
    &= \frac{2\omega_o^2}{\omega_q^2 + (\Gamma/2)^2} \Bigg\{ A_\theta \left[ e^{-\Gamma |\xi|/2} \frac{(\omega_q^2 - (\Gamma/2)^2)\cos\omega_q\xi + \Gamma \omega_q \sin\omega_q|\xi|}{\omega_q^2 + (\Gamma/2)^2} - \Gamma|\xi| \right] \nonumber \\
    &~~~~~~~~~~~~~~~~~~~ - B_\theta \left[e^{-\Gamma \eta} \frac{(\omega_q^2 - (\Gamma/2)^2)\cos2(\omega_q \eta + \varphi) + \Gamma \omega_q \sin2(\omega_q \eta + \varphi)}{\omega_q^2 + (\Gamma/2)^2} - \Gamma\left( \frac{\omega_q \sin2\varphi}{\omega_q^2 + (\Gamma/2)^2} + \eta \cos2\varphi \right) \right] \Bigg\}, \nonumber
\end{align}
where $\eta > 0$.
Moreover, by considering our approximation, $\omega_q\gg \Gamma$, we have
\begin{align}
    \omega_o^2 \int_{-\infty}^{\infty} d\omega~J(\omega)\omega^{-2} \big[ A_\theta \cos\omega\xi - B_\theta \cos2(\eta+\varphi) \big] 
    \approx \bigg(\frac{\omega_o}{\omega_q}\bigg)^2 \Bigg\{ A_\theta  e^{-\Gamma |\xi|/2} \cos\omega_q\xi - B_\theta e^{-\Gamma \eta} \cos2(\omega_q \eta + \varphi)  \Bigg\} \label{eq:term3_limit}
\end{align}

We finally have Eq.\eqref{eq:all_terms} just by summation of Eqs.\eqref{eq:term1_final}, \eqref{eq:term2_final} and \eqref{eq:term3_final}. Combining all terms together and assuming $\omega_q \gg \Gamma$, we have the regularized noise kernel
\begin{equation}
     \langle\eta(t)\eta(t')\rangle_0 \approx \left[\frac{1+\cos^2\theta}{2}(1+r^2)+2r\right] e^{-\frac{\Gamma}{2}|t-t'|}\cos(t-t') - \frac{\sin^2\theta}{2}(1+r)^2 e^{-\frac{\Gamma}{2}(t+t')}\cos(t+t'+2\varphi)
\end{equation}
and by the same reasoning, we have the deterministic force
\begin{equation}
    f_0(t; \theta, \varphi) \approx e^{-\frac{\Gamma}{2} t}\cos(t+\varphi).
\end{equation}

\end{document}